\documentclass[aps,pre,twocolumn,groupedaddress]{revtex4-1}

\usepackage{graphicx,bbm,bm,hyperref}  
\usepackage{amsmath,amssymb}
\usepackage{amsthm}

\usepackage[pdftex]{color}

\def\ii{{\rm i}}

\def\Z2{\mathbbm{Z}_2}

\def\lph{\lambda_{\rm ph}}
\def\leff{\lambda_{\rm eff}}
\def\bI{\mathbf{I}}
\def\Aker{A_{\rm ker}}
\def\R{{\mathbf{R}}}
\def\L{{\mathbf{L}}}
\def\r{{\mathbf{r}}}
\def\l{{\mathbf{l}}}
\def\ra#1{{\tilde{\mathbf{r}}_{#1}}}
\def\la#1{{\tilde{\mathbf{l}}_{#1}}}
\def\raa#1{{\tilde{\mathbf{r}}^{(1)}_{#1}}}
\def\laa#1{{\tilde{\mathbf{l}}^{(1)}_{#1}}}
\def\tK{{t_{\rm K}}}
\def\tc{{t_{\rm c}}}
\def\tR{\tilde{\mathbf{R}}}
\def\tL{\tilde{\mathbf{L}}}

\def\tr#1{{\rm tr}(#1)}
\def\1{\mathbbm{1}}
\def\ket#1{{| #1 \rangle}}
\def\bra#1{{\langle #1 |}}
\def\braket#1#2{{\langle #1 | #2 \rangle}}
\def\bracket#1#2#3{{\langle #1|#2 | #3 \rangle}}

\def\tit#1{{\em #1},}

\begin{document}

\title{Phantom relaxation rate of the average purity evolution in random circuits due to Jordan non-Hermitian skin effect and magic sums}

\author{Marko \v Znidari\v c}
\affiliation{Physics Department, Faculty of Mathematics and Physics, University of Ljubljana, 1000 Ljubljana, Slovenia}

\date{\today}

\begin{abstract}
Phantom relaxation is relaxation with a rate that is not given by a finite spectral gap. Studying the average purity dynamics in a staircase random Haar circuit and the spectral decomposition of a non-symmetric matrix describing the underlying Markovian evolution, we explain how that can arise out of an ordinary-looking spectrum. Crucial are alternating expansion coefficients that diverge in the thermodynamic limit due to the non-Hermitian skin effect in the matrix describing the average purity dynamics under an overall unitary evolution. The mysterious phantom relaxation emerges out of localized generalized eigenvectors describing the Jordan normal form kernel, and, independently, also out of interesting trigonometric sums due to localized true eigenvectors. All this shows that when dealing with non-Hermitian matrices it can happen that the spectrum is not the relevant object; rather, it is the pseudospectrum, or, equivalently, a delicate cancellation enabled by localized eigenvectors.
\end{abstract}

\maketitle

\section{Introduction}

Quantum evolution under Hermitian generators evolves states via unitary propagators, thereby preserving the norm and the inner products between states. Similarly, classical evolution in phase space is governed by a unitary Koopman propagator~\cite{Braun}. Our physics intuition is therefore often built upon properties of Hermitian/unitary matrices. An example is a focus on the real spectrum of Hermitian matrices, for instance in low-temperature condensed matter~\cite{sachdev}, or using various spectral objects as indicators of quantum chaos~\cite{Haake}. However, real systems are never really closed, naturally leading to non-Hermiticity, an example being the Lindblad master equation~\cite{breuer}. Perhaps even more importantly, in a many-body system we can never control and observe all exponentially many observables. In such a situation one invariably either traces out some degrees of freedom, or does a coarse graining, again ending up with a non-Hermitian description. A paradigmatic example is the II.~law of thermodynamics where the ``arrow of time'' arises out of unitary dynamics.

The question one can then ask is how good our intuition gained from Hermitian linear algebra works on non-Hermitian matrices? In the present work we shall demonstrate that it can be completely wrong. In particular, the spectrum of a non-Hermitian matrix is not necessarily the object determining physics -- rather, it is the eigenvectors. Because these eigenvectors need not be orthogonal they can lead to large spectral expansion coefficients, in turn resulting in a delicate cancellation leading to slow phantom relaxation (which would be impossible according to Hermitian intuition). Another twist can be added by a non-diagonalizable extensively large Jordan normal form block, i.e. an exceptional point, with localized generalized eigenvectors.

Let us consider a process whose dynamics is described by a matrix iteration, $\bI(t+1)=A\, \bI(t)$, where $\bI(t)$ is some state vector at time $t$ and $A$ is a time-independent $n \times n$ matrix. Let the matrix $A$ have the largest eigenvalue $\lambda_0=1$, with the corresponding right eigenvector $\mathbf{I}_\infty=\bI(\infty)$ describing a steady state, $A \mathbf{I}_\infty=\mathbf{I}_\infty$, while all other eigenvalues $\lambda_j$ are strictly smaller in modulus, $|\lambda_j|<1$. One would expect that if the gap $1-|\lambda_1|$ is finite in the thermodynamic limit (TDL) $n \to \infty$, then the relaxation rate towards the steady state $\bI(t)-\mathbf{I}_\infty$ will be given by that 2nd largest eigenvalue $\lambda_1$.

We explain why such a statement can be wrong if $A$ is non-Hermitian (non-symmetric), and how a so-called phantom relaxation $|\bI(t)-\mathbf{I}_\infty| \sim \lph^t$, where the rate is given by a non-existent phantom ``eigenvalue'' $\lph$ such that $|\lambda_1| < \lph < 1$, can arise out of spectral decomposition in the TDL, i.e., for a many-body system. The phenomenon has been observed for purity in several previous works~\cite{PRX21,PRR,PRA23}, as well as in out-of-time-ordered correlation (OTOC) functions~\cite{PRR22}, and in non-random Floquet systems~\cite{arxiv}. While our explicit example is for $A$ being the average purity propagator, similar mathematical properties should be possible in other situations involving non-Hermitian matrices, for instance transfer matrices~\cite{lamacraft21,pieter23}. Indeed, it has been observed that the gap does not necessarily give the correct relaxation rate in Lindblad master equations~\cite{song,takashi20,takashi21,ueda21,Mori23}, and that the emergence of true relaxation rate can be due to many contributing eigencomponents~\cite{clerk}.

Focusing on an example where $A$ describes the average purity evolution under staircase random circuit we use the spectral decomposition of $A$ to show how $\lph$, that is not visible anywhere in the spectrum, somehow magically emerges out of localized left and right eigenvectors, also known as a non-Hermitian skin effect~\cite{Wang18,Kunst21,Ueda20,Torres20,Okuma22,Lee23}. At short but extensive times one also needs to take into account an extensively large Jordan normal form kernel, which also shows a non-Hermitian skin effect with localized generalized eigenvectors. We stress that the non-Hermitian skin effect in our case does not arise out of an explicit non-Hermitian generator, or an outside noise, but emerges naturally out of a non-symmetric matrix that describes the reduced dynamics after tracing out part of the system. The Jordan kernel is also instrumental for a rapid emergence of the pseudospectrum, i.e., the spectrum of a slightly perturbed matrix, which is an alternative way to get $\lph$~\cite{PRR}. There are other recent observations that the pseudospectrum has physical relevance~\cite{Okuma20b,Okuma21,Viola21,Yoshida21}.

\subsection{Summary of results}

We study purity evolution under unitary random circuit starting with an $n$-qudit product initial state $\ket{\psi(0)}$, with local qudit Hilbert space dimension denoted by $d$. The state at time $t$ is $\ket{\psi(t)}=U(t)\ket{\psi(0)}$, where $U(t)$ is a unitary that is a product of random 2-qudit gates (explicit details will be given later). The purity $I_k(t)$ that measures bipartite entanglement between the first $k$ and the last $n-k$ qudits is
\begin{equation}
  I_k(t)=\tr{\rho_k^2(t)},\quad \rho_k(t)={\rm tr}_{k+1,\ldots,n}\ket{\psi(t)}\bra{\psi(t)}.
  \label{eq:Ik1}
\end{equation}
The purity is initially $I_k(0)=1$ and decays towards its random state value~\cite{Lubkin} $[\mathbf{I}_\infty]_k=I_k(t \to \infty)$ at long times. Note that the purity of the subsystem, i.e., of the first $k$ qudits, decreases because we are tracing out a part of the system -- the evolution on the whole system of $n$ qudits is unitary. Under such unitary evolution the purity measures genuine bipartite entanglement~\cite{footX}. Relaxation of purity is then studied by looking at $\Delta I_k(t)$,
\begin{equation}
  \Delta I_k(t)=I_k(t)-[\mathbf{I}_\infty]_k,\qquad [\mathbf{I}_\infty]_k=\frac{d^k+d^{n-k}}{1+d^n},
  \label{eq:Iinf}
\end{equation}
where we use square brackets to denote a component of a (boldface) vector. One can show that the average evolution of quantities quadratic in the evolved state is Markovian~\cite{Oliveira}. Compactly writing purities $I_k(t)$ for all $k$ as components of a vector~\cite{foot1} $\bI$, evolution of average purities (averaged over random circuit randomness) is governed by a simple matrix iteration~\cite{Kuo20},
\begin{equation}
  \bI(t+1)=A\,\bI(t).
\end{equation}
There are several equivalent ways of writing down such a matrix iteration~\cite{Oliveira,PRA08,Kuo20,PRR}; in the most compact one~\cite{PRR}, that we also use, $A$ is an $n\times n$ dimensional real matrix.
\begin{figure}
  \centerline{\includegraphics[width=.9\columnwidth]{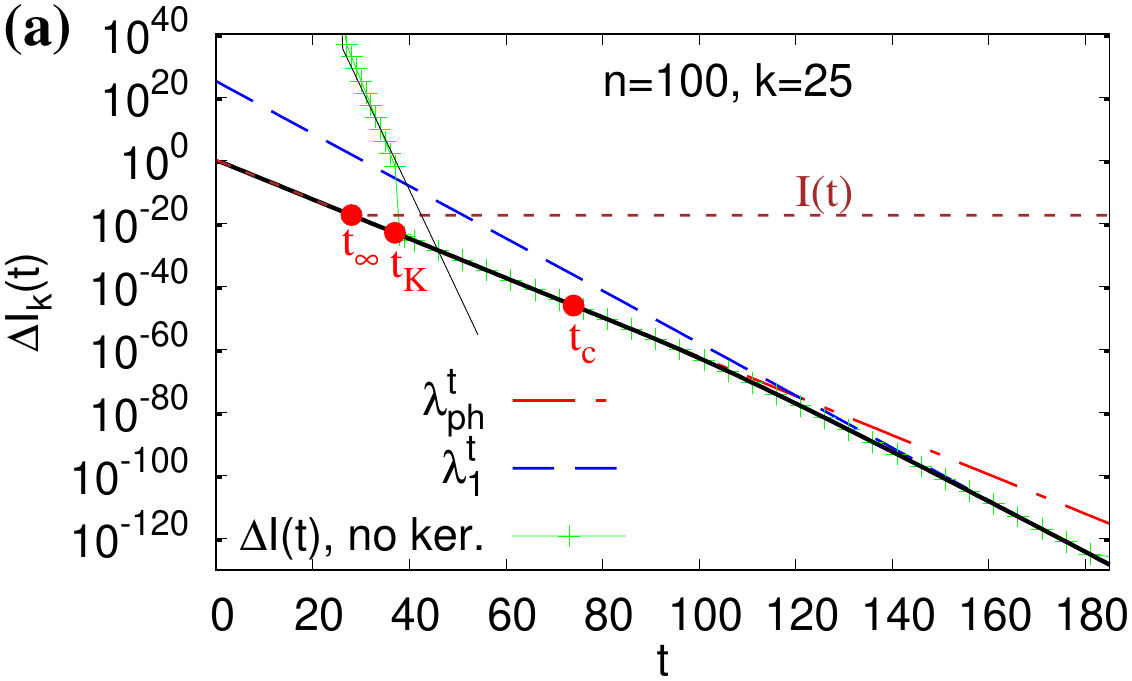}}
  \vskip2pt
\centerline{\includegraphics[width=.59\columnwidth]{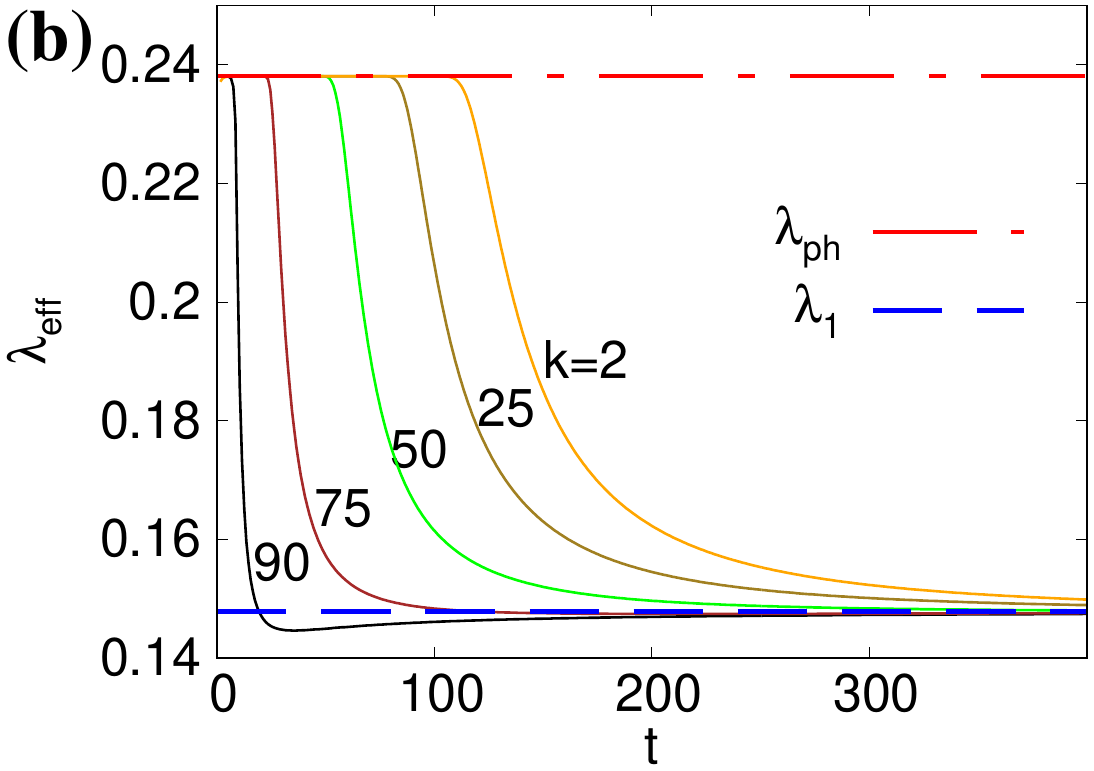}\hskip1pt\includegraphics[width=.43\columnwidth]{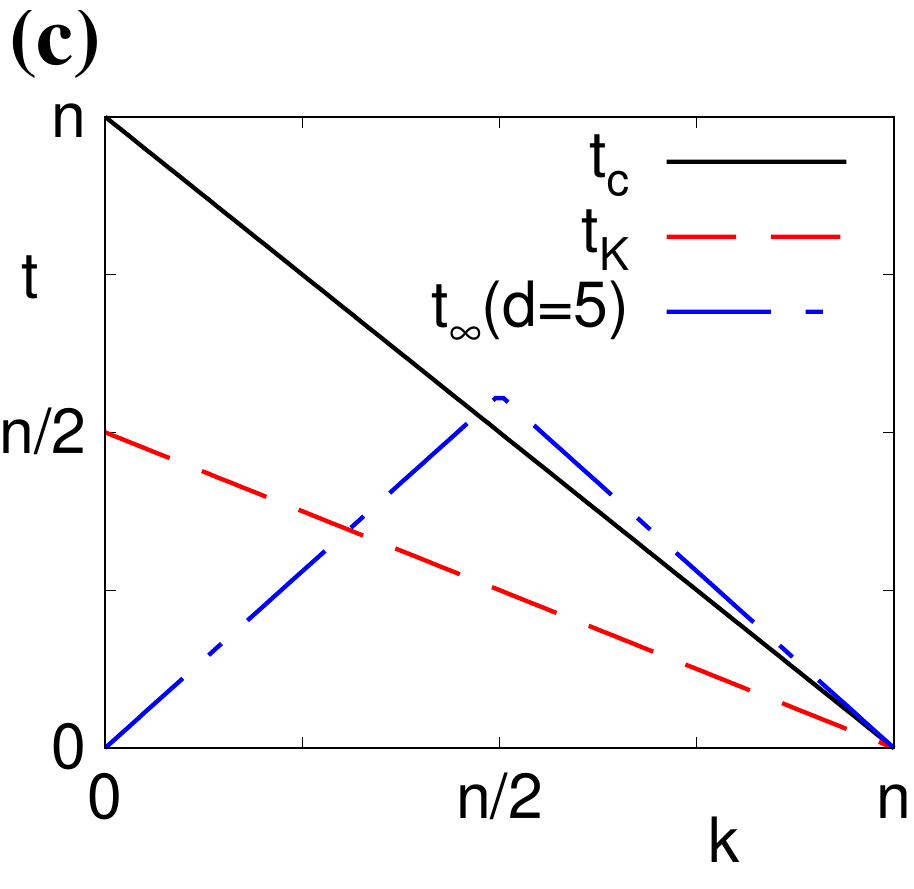}}
\caption{Relaxation of purity to its asymptotic value. (a) Full black line shows exact $\Delta I_k(t)$, which initially decays with a phantom rate as $\lph^t$ (red chain line), after which it eventually transitions into $\lambda_1^t$ (blue dashed line). Brown curve that saturates shows $I_k(t)$. Red points mark theoretical purity saturation time $t_\infty$ (\ref{eq:tinf}), time $t_{\rm K}$ (\ref{eq:tK}) until the kernel contributes, and $t_{\rm c}$ (\ref{eq:tK}) when the effective rate starts to deviate from $\lph$. Green pluses show purity evolution neglecting the kernel (full thin black line overlapping with diverging green pluses is Eq.(\ref{eq:diver})). (b) Instantaneous relaxation rate $\leff$ (defined via $\Delta I_k(t) \sim \leff^t$ \ref{eq:leff}), from which one can more clearly see the transition, for the same $n=100$ and $k=25$ as in (a), as well as for bipartitions with $k=2,50,75,90$. (c) Dependence of three timescales marked in (a) on the bipartition index $k$.}
\label{fig:I100}
\end{figure}

Instead of purity one often studies also the 2nd R\' eny entropy $S_2(t)=-\ln I_k(t)$. In principle $S_2$ can behave differently than the purity (the average of the logarithm vs. the logarithm of the average), however, for our specific Haar random circuit one would observe the phantom relaxation also in the 2nd R\' eny entropy, as well as in other R\' eny entropies; for numerical data in slightly different circuits see Refs.~\cite{PRX21,arxiv}.

We are going to use the spectral decomposition of $A$. In fact, $A$ has an $\frac{n}{2}-1$ dimensional kernel $\Aker$ composed of a single Jordan block, so that we can write
\begin{equation}
  A=A_\lambda+\Aker,
  \label{eq:Alamker}
\end{equation}
where $A_\lambda$ is diagonalizable and contains all nonzero eigenvalues, and can be spectrally decomposed as
\begin{equation}
  A_\lambda=\sum_j \lambda_j \ket{\R_j}\bra{\L_j},
  \label{eq:Aspec}
\end{equation}
with biorthogonal left and right eigenvectors. The kernel $\Aker$ contributes to the purity $I_k(t)$ only for $t \le \tK$, where
\begin{equation}
  t_{\rm K}=\left\lfloor \frac{t_{\rm c}}{2} \right\rfloor,\qquad   \tc=n-k-1.
  \label{eq:tK}
\end{equation}
and can therefore be ignored for $t>\tK$. Writing out localized left and right eigenvectors of $A_\lambda$ (\ref{eq:Aspec}), projecting the initial vector of purities $\bI(0)$ on the left eigenvectors, one gets a spectral resolution of $\Delta I_k(t)$, 
\begin{equation}
  \Delta I_k(t)=\sum_{j=1}^{n/2-1} c_j \lambda_j^t,\qquad \lambda_j=(2\alpha)^2 \cos^2{\left( \frac{\pi j}{n}\right)},
  \label{eq:sumcj}
\end{equation}
in terms of non-steady-state eigenvalues $\lambda_j$ that are all non-degenerate, and expansion coefficients $c_j$, and $\alpha=d/(d^2+1)$. In particular, the largest eigenvalue in the sum is in the TDL
\begin{equation}
  \lambda_1=(2\alpha)^2.
  \label{eq:lambda1}
\end{equation}
Using special properties of certain (magic) trigonometric sums involved in $c_j$ we show that one has an exact phantom decay $\Delta I_k(t)=\lph^t$ for $t<\tc$ (\ref{eq:tK}), with
\begin{equation}
  \lph=\frac{\alpha}{1-\alpha}=\sum_{k=1}^\infty \alpha^k.
\end{equation}
In Fig.~\ref{fig:I100}(a) we show an example of purity decay, illustrating phantom decay for $t<\tc$. As long as $n-k$ is extensive ($n-k \propto n$) this time grows with the system size, and therefore $\lph$, and not $\lambda_1$, is the relevant decay in the TDL~\cite{foot1a}. We find that $\lph$, instead of $\lambda_1$, comes into play due to $\bI(0)$ having an extensive number of nonzero components, i.e., a nonzero overlap with localized $\ket{\L_j}$. 

\begin{figure}
  \centerline{\includegraphics[width=.8\columnwidth]{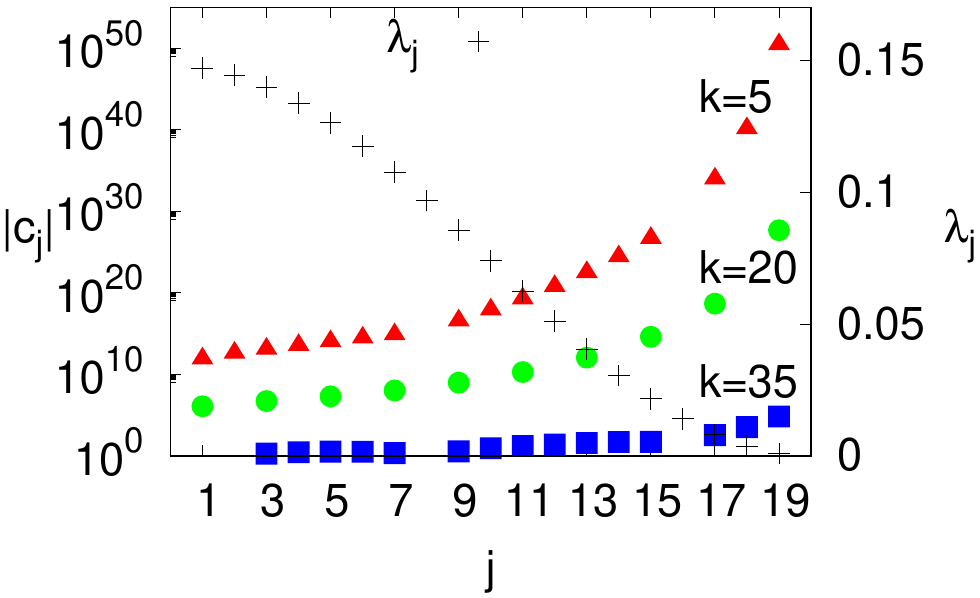}}
  \caption{Absolute values of eigenvalue expansion coefficients $c_j$ of purity $I_k$ (\ref{eq:sumcj}). They alternate in signs and grow rapidly with eigenvalue index $j$. Shown is data for $n=40$ and bipartitions with $k=5,20,35$ (full red, green, blue points). With black pluses are shown the corresponding eigenvalues $\lambda_j$ (right axis). Missing points, e.g. at even $j$ for $k=20$, or $j=8,16$ for $k=5,35$, are where $c_j=0$.}
  \label{fig:cj}
\end{figure}

Because $0<\lambda_j<1$, the only way the slower decay with $\lph>\lambda_1$ can emerge is if $c_j$ (\ref{eq:sumcj}) get large with $n$, and if they alternate in sign ($c_j$ are real). Namely, if $c_j$ would be of the same sign one could bound $\Delta I_k(t+1) \le \Delta I_k(t) \lambda_1$ because $c_j \lambda_j^{t+1} \le c_j \lambda_j^t \lambda_1$, and therefore the decay could not be slower than $\lambda_1$. The origin of the slower phantom decay is in a delicate cancellation of many diverging expansion coefficients $c_j$, see Fig.~\ref{fig:cj}. In addition to growing with $n$ they also grow rapidly with the eigenvalue index $j$, i.e., $c_j$ corresponding to smaller eigenvalues $\lambda_j$ are much larger than those corresponding to larger ones. This means that, contrary to intuition, in the spectral sum (\ref{eq:sumcj}) one cannot neglect smaller $\lambda_j$. In fact, quite the opposite. Up-to some time (that can scale as $\propto n$) terms with smaller $\lambda_j$ give in modulus a larger contribution. All this is due to a non-Hermitian skin effect, i.e., exponential localization of eigenvectors of $A_\lambda$, causing the expansion coefficients $c_j$ to blow up with $n$ (and $j$).

For $t>\tc$ the effective decay $\leff$ defined by $\Delta I_k(t) \sim \leff^t$ starts to transition towards $\lambda_1$. The transition is not sharp, see Fig.~\ref{fig:I100}(b), with the shape being described in the TDL by Jacobi theta functions. Using appropriate asymptotics we show that $\leff(t)-\lambda_1$ reaches any $n$-independent value at a time scaling as $\sim n$, while it becomes small of the order $\sim 1/n^2$ only at a longer time scaling as $\sim n^2$.

The kernel can be written in terms of generalized eigenvectors~\cite{Weintraub} as
\begin{equation}
  \Aker=\sum_{k=1}^{n/2-2} \ket{r_k}\bra{l_{k+1}},
  \label{eq:Aker}
\end{equation}
where $\braket{r_k}{l_j}=\delta_{k,j}$. Due to the shift property of generalized eigenvectors, $A \ket{r_k}=\ket{r_{k-1}}$, and $A^{\rm T} \ket{l_k}=\ket{l_{k+1}}$, we can also immediately write powers,
\begin{equation}
  A_{\rm ker}^t=\sum_{k=1}^{n/2-t-1} \ket{r_k}\bra{l_{k+t}}.
  \label{eq:Akert1}
\end{equation}
As mentioned, the kernel $\Aker$ is important only for $t \le \tK$ (\ref{eq:tK}) where however $\Aker$ is crucial to get the correct phantom decay. Despite no explicit time-dependence, except in the size of the kernel and vector pairings (\ref{eq:Akert1}), writing down the contribution of $\Aker$ to $\Delta I_k(t)$ we find that this gets reflected in the time dependence that is such as to exactly cancel a diverging contribution from $A_\lambda$ (green pluses for $t\le t_{\rm K}$ in Fig.~\ref{fig:I100}(a)). In a way, similarly to $t > \tK$, where localization of eigenvectors was crucial, to get the slow decay with $\lph>\lambda_1$ for $t\le \tK$ it is the non-Hermitian skin effect (localization) of generalized eigenvectors that is important.

Therefore, we learn that in non-Hermitian matrices it can happen that the spectrum is not the relevant object determining relaxation. Rather, it is the localization (non-Hermitian skin effects) of eigenvectors, or, as we find for $t \le \tK$, also of generalized eigenvectors of the Jordan normal form structure of the kernel.

It has been shown in Ref.~\cite{PRR} that $\lph$ is equal to the norm of the pseudospectrum -- the spectrum of a slightly perturbed matrix $A$. While the pseudospectrum is independent of the perturbation strength $\varepsilon$ in the TDL~\cite{trefethen}, we show that the speed with which it is established in a finite system crucially depends on the extensive size of the Jordan kernel. Namely, the effect of perturbation scales in a non-linear way as $\varepsilon^{1/(n/2-1)}$, meaning that in the TDL limit even exponentially small perturbation will completely change the spectrum of $A$. The spectrum of $A$ is therefore very sensitive, while the pseudospectrum is a robust object~\cite{trefethen}.

\section{Purity decay and its matrix description}

\begin{figure}[t]
\centerline{\includegraphics[width=.6\columnwidth]{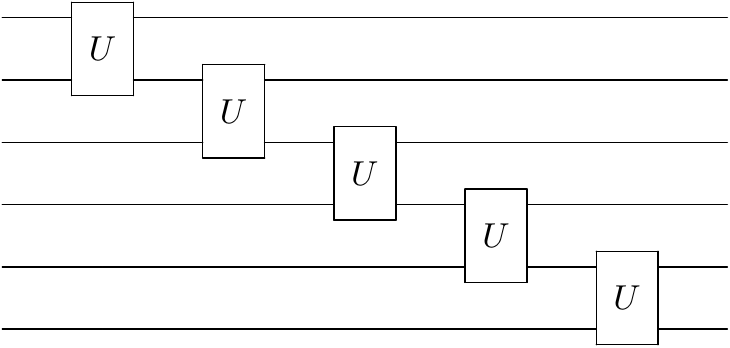}}
\caption{Staircase configuration of gates on $n=6$ qudits. Time runs from left to right, and the 1st qudit is the top-most. All nearest-neighbor gates $U$ are random independent Haar distributed unitaries.}
\label{fig:S}
\end{figure}

We are interested in the purity evolution under a random circuit. The random circuit we shall study is composed of random independent two-qudit gates drawn according to unitarily invariant Haar measure~\cite{Karol}. The unitary propagator for one unit of time $U(1)$ is a product of all nearest-neighbor gates applied in a staircase order (Fig.~\ref{fig:S}),
\begin{equation}
  U(1)=U_{n,n-1} U_{n-1,n-2} \cdots U_{1,2},
\end{equation}
where $U_{i,i+1}$ are all independent and distributed according to the Haar measure, in other words, they are random $U(d^2)$ unitaries acting on $d$-dimensional local space of qudits. The unitary propagator for time $t$ is a product of $t$ such one-step propagator $U(1)$, with all gates being independent and identically distributed. Random circuits have a long history, going back to early~\cite{Lloyd} and more recent experiments~\cite{google,Liu23}, as well as numerous theoretical works exploring their statistical and dynamical properties, including Refs.~\cite{Harrow09,Adam18,Frank18,Chan18,vedika,Zhou19,Nick22,Zhou22}.

The dynamics of the average purity is a Markovian process~\cite{Oliveira,PRA08,Kuo20} allowing one to write the time evolution of the average purity for all $2^n$ different bipartitions as a simple matrix iteration. The matrix involved is exponentially large in $n$, however, for the staircase circuit with open boundary conditions and bipartitions with a single boundary (single-cut) between the two subsystems, i.e., $I_k$ denotes purity for a bipartition into first $k$ qudits and the rest, one can write the evolution for those bipartitions in terms of a matrix of size $n$~\cite{PRR}. Defining an $n$-component vector of purities $\mathbf{I}=(1,I_2,I_3,\ldots,I_{n-1},1)$~\cite{foot2} one has~\cite{PRR}
\begin{equation}
  \mathbf{I}(t+1)=A \bI(t),
  \label{eq:Iiter}
\end{equation}
where the $n\times n$ matrix $A$ can be written in terms of two $(n-2)$ dimensional column vectors $\mathbf{a}_{1,2}$, and a $(n-2)$ dimensional square Toeplitz matrix $T$,
\begin{equation}
  A=\begin{pmatrix}
    1 & 0 & 0 \\
   \mathbf{a}_1 & T & \mathbf{a}_2 \\
   0 & 0 & 1
\end{pmatrix},\qquad
\begin{array}{rcl}
  \mathbf{a}_1&=&(\alpha^2,\alpha^3,\ldots,\alpha^{n-1})\\
  \mathbf{a}_2&=&(0,\ldots,0,\alpha)
\end{array}
\label{eq:Am}
\end{equation}
where $T$ is
\begin{equation}
T=\begin{pmatrix}
    \alpha^2 & \alpha & 0 & \cdots & 0 \\
    \alpha^3 & \alpha^2 & \alpha & \cdots & 0\\
    \vdots   &  \vdots & \ddots & \cdots & \vdots \\
    \alpha^{n-2} & \alpha^{n-3} & \ddots & \ddots & \alpha\\
    \alpha^{n-1} & \alpha^{n-2} & \cdots & \cdots & \alpha^2 
\end{pmatrix}.
\label{eq:Ts}
\end{equation}
Dependence on the local Hilbert space dimension $d$ is contained in the only parameter $\alpha$,
\begin{equation}
  \alpha=\frac{d}{d^2+1}.
  \label{eq:alpha}
\end{equation}
All the interesting properties that we discuss will follow from this simple matrix $T$. Its most important property is that it is not symmetric which will lead to a non-Hermitian skin effect. Note that this non-Hermiticity is not put in ``by hand'' like in many non-Hermitian skin effect studies, but emerges automatically from the underlying unitary evolution.

Simplification of the average purity evolution from an exponentially large matrix to one that is only polynomial in $n$ is possible also for multi-cut bipartitions~\cite{PRA23}, with the matrix though being much more complicated. In this work we stick to the single-cut bipartitions where the simple form (\ref{eq:Am}) will allow for a full spectral decomposition. We should also mention that while the exact matrix iteration (\ref{eq:Iiter}) results only after averaging over Haar random gates, the phenomenon of phantom relaxation is due to self-averaging observed also on a single-realization basis and for other R\' eny entropies~\cite{PRX21}, as well as in some Floquet systems without any randomness~\cite{arxiv}.

Starting with a fully separable initial state all initial purities are $1$, and therefore the initial vector is
\begin{equation}
  \bI(0)=(1,\ldots,1).
  \label{eq:I0}
\end{equation}
Purity for bipartition $k$ is then simply,
\begin{equation}
  I_k(t)=\braket{\mathbf{e}_k}{A^t \bI(0)},
  \label{eq:Ik}
\end{equation}
where $\mathbf{e}_k$ is the $k$-th basis vector, i.e., $[\mathbf{e}_k]_p=\delta_{p,k}$. Iteration in Eq.(\ref{eq:Iiter}) with matrix $A$ (\ref{eq:Am}) is the starting point of our work. For details on its derivation see Ref.~\cite{PRR}.

Because we want to focus on the relaxation process towards a state reached at long times, and in a finite system purity at long times is not zero, we shall subtract from $I_k(t)$ its long-time random state value $I_k(\infty)$ and study $\Delta I_k(t)$ (\ref{eq:Iinf}). In Ref.~\cite{PRR} it has been shown by directly solving recursive relations that the initial decay is $\Delta I_k(t) = \lph^t$ in the leading order in $n$, with the phantom decay rate given by
\begin{equation}
\lph=\frac{\alpha}{1-\alpha}=\sum_{k=1}^\infty \alpha^k.
\end{equation}
Independently, it has been also observed~\cite{PRR} that this $\lph$ is exactly equal to the norm of the pseudospectrum of $A$, which can be exactly calculated for Toeplitz matrices (the psuedospectrum is given by the so-called symbol of the Toeplitz matrix~\cite{trefethen}, which in physics language is nothing but the Bloch Hamiltonian, i.e., Fourier transformation of values along the antidiagonal). 

Knowing $\lph$ we can estimate the time $t_\infty$ when the purity becomes close to $I_k(\infty)$, see Fig.~\ref{fig:I100}. Equating $\lph^{t_\infty}=I_k(\infty)$ gives
\begin{equation}
  t_\infty={\rm min}(k,n-k) \frac{\ln{d}}{-\ln{\lph}}.
  \label{eq:tinf}
\end{equation}
Before diving into spectral decomposition of $A$ let us comment on the form of $A$. The main crux will be obtaining and handling spectral properties of the matrix $T$. Namely, eigenvectors of $A$ will be readily constructed from those of $T$, while the spectrum of $A$ is a union of the spectrum of $T$ and two steady-state eigenvalues $\lambda_0=1$.

Matrix $T$ is rather interesting: it is one of the simplest matrices that has an extensively large kernel and at the same time a nontrivial spectrum and eigenvectors. Some simple deformations of $T$ result in simpler spectral properties, e.g., (i) setting the superdiagonal to $0$ results in a lower triangular matrix that has only a $(n-2)$ dimensional Jordan block corresponding to the eigenvalue $\alpha^2$ (and no nonzero eigenvalues), (ii) deleting the lower triangle, i.e., keeping the diagonal and superdiagonal, would result in the same, and (iii) keeping only the superdiagonal would again result in a single Jordan block with eigenvalue $0$. Knowing the spectral properties of $T$ could therefore be of interest in itself in linear algebra, for instance for analysis of numerical algorithms, where special badly conditioned matrices are used as test cases for numerical linear algebra algorithms~\cite{matrixmarket}. In fact, replacing the first column in $T$ with zeros and using $\alpha=\frac{1}{2}$ gives the iteration matrix of a Gauss-Seidel iterative method of solving a set of linear equations~\cite{trefethen}, where the speed of the algorithm directly relates to the pseudospectrum of the matrix. Another case is taking $-\alpha$ instead of $\alpha$ in the superdiagonal and setting $\alpha=1$, which results in a so-called Grcar matrix~\cite{grcar}.

\section{Spectral decomposition}

In this section we are going to write the spectral decomposition of $A_\lambda$ and get the corresponding spectral sum in Eq.(\ref{eq:sumcj}). This will get us correct purity relaxation for $t>\tK$ (\ref{eq:tK}), including $\lph$ and the transition behavior from $\lph$ to $\lambda_1$. In the subsequent section we are then going to study $\Aker$, which will also explain the behavior for $t \le \tK$.

$A_\lambda$ can be decomposed as
\begin{equation}
  A_\lambda=\ket{\R_0}\bra{\L_0}+\ket{\R_0'}\bra{\L_0'}+\sum_{j=1}^{n/2-1} \lambda_j \ket{\R_j}\bra{\L_j},
  \label{eq:Alam}
\end{equation}
where we use the bra/ket notation for left/right eigenvectors that satisfy biorthogonality
\begin{equation}
  \braket{\L_k}{\R_p}=\delta_{k,p}.
  \label{eq:orth}
\end{equation}
In addition to biorthogonality (\ref{eq:orth}) one can also enforce normalization of one of them (but not both), e.g., $\braket{\R_k}{\R_k}=1$, or $\braket{\L_k}{\L_k}=1$. The first two projectors in Eq.(\ref{eq:Alam}) correspond to two steady states, where as we will see only one is physically relevant.

We shall first write down eigenvectors of $T$, and then due to simple connection between $T$ and $A$ also those of $A$. The left and right eigenvectors and $\lambda_j$ of $T$ have already been obtained in Refs.~\cite{PRR,Fairweather71}; we write them down again for completeness. The eigenvalues of $T$ are
\begin{equation}
  \lambda_j=\left(2\alpha \cos\varphi_j\right)^2,\quad \varphi_j=\frac{\pi j}{n},\quad j=1,\ldots, \frac{n}{2}-1.
  \label{eq:lambdaj}
\end{equation}
Throughout the paper we for simplicity assume that $n$ is even so that $n/2$ is an integer. Denoting by $\tL_j$ and $\tR_j$ the corresponding unnormalized (not satisfying Eq.(\ref{eq:orth})) eigenvectors of $T$, their components are
\begin{eqnarray}
[\tR_j]_k &=& (2\alpha \cos{\varphi_j})^{k-2} \frac{\sin{[(k+1)\varphi_j]}}{\sin{\varphi_j}}, \\
{[}\tL_j]_k &=& (2\alpha \cos{\varphi_j})^{n-3-k} \frac{\sin{[(n-k)\varphi_j]}}{\sin{\varphi_j}},
  \label{eq:tLR}
\end{eqnarray}
where $k=1,\ldots,n-2$, and $\varphi_j$ used throughout the paper is given in Eq.(\ref{eq:lambdaj}). Note that left eigenvectors are simply the reflected right ones, i.e., can be obtained by replacing the component index $k$ by $n-k-1$. Right eigenvectors are exponentially localized at the left edge (small indices $k$), while the left are localized at the right edge, with the localization length being $-1/\ln{(2\alpha\cos{\varphi_j})}$ which is upper-bounded by an $n$-independent $-1/\ln{(2\alpha)}$. The inner product between left and right eigenvectors can be simplified using $2\sin{\alpha}\sin{\beta}=\cos{(\alpha-\beta)}-\cos{(\alpha+\beta)}$, resulting in
\begin{equation}
  N_j=\sum_{k=1}^{n-2} [\tR_j]_k [\tL_j]_k=(-1)^{j+1} (2\alpha \cos{\varphi_j})^{n-5} \frac{n\cos{\varphi_j}}{2\sin^2{\varphi_j}}.
  \label{eq:Nj}
\end{equation}
This normalization is exponentially small in $n$ (remember that (\ref{eq:alpha}) implies $\alpha<1$ for $d>1$). When enforcing biorthogonality (\ref{eq:orth}), which means we will have do divide one of them by $N_j$, this will cause exponentially large expansion coefficients in the spectral decomposition.

From eigenvectors of $T$ we can construct those of $A$,
\begin{eqnarray}
  \R_j&=&\frac{1}{N_j}(0,\tR_j,0),\nonumber \\
  \L_j&=&( \frac{\braket{\tL_j}{\mathbf{a}_1}}{\lambda_j-1},\tL_j,\frac{\braket{\tL_j}{\mathbf{a}_2}}{\lambda_j-1}),
  \label{eq:LR}
\end{eqnarray}
that is, components $k=2,\ldots,n-1$ of $\L_j$ and $\R_j$ are exactly equal to $n-2$ components of $\tL_j$ and $\tR_j$, respectively, while the 1st and the last one are as written in Eq.(\ref{eq:LR}). Note that those eigenvectors already satisfy biorthogonality (\ref{eq:orth}) due to dividing $\tR_j$ by $N_j$. Two steady-states (\ref{eq:Alam}) corresponding to $\lambda_0=1$ are
\begin{eqnarray}
  \R_0&=& (1,\mathbf{I}_\infty,1), \nonumber \\
  \L_0&=& (\frac{1}{2},0,\ldots,0,\frac{1}{2}),
  \label{eq:SS}
\end{eqnarray}
where $\mathbf{I}_\infty$ is a vector of purity saturation values (\ref{eq:Iinf}) for $k=2,\ldots,n-1$. The 2nd steady-state is
\begin{eqnarray}
  \L_0'&=& (\frac{1}{2},0,\ldots,0,-\frac{1}{2}),
  \label{eq:SS2}
\end{eqnarray}
while the right one $\R_0'$, which we do not need, is given in Ref.~\cite{foot4}. Because the purity vector $\bI(t)$ (\ref{eq:Iiter}) always has the 1st and last component equal to $1$, including at $t=0$ (and irrespective of the initial state being fully separable), one has $\braket{\L_0'}{\bI(0)}=0$ and the 2nd steady-state (\ref{eq:SS2}) does not contribute to purity evolution. The steady state $\R_0$ (\ref{eq:SS}) gives exactly the asymptotic purity (as it should be), and therefore the contribution of nonzero eigenvalues, i.e. $A_\lambda$, to purity relaxation is given by Eq.(\ref{eq:sumcj}) with the expansion coefficients,
\begin{equation}
  c_j=\braket{\mathbf{e}_k}{\R_j}\, \braket{\L_j}{\bI(0)}=[\R_j]_k \sum_{p=1}^n [\L_j]_p,
  \label{eq:cj}
\end{equation}
where we plugged in $\bI(0)=(1,1,\ldots,1)$ for a fully separable initial state.

\subsection{Expansion coefficients}

As we have explained a necessary condition to observe phantom decay is that $c_j$ are not all of the same sign. From Eqs.(\ref{eq:cj}) and (\ref{eq:LR}) we immediately see that the alternating sign of $c_j$ will come from normalization $N_j$ (\ref{eq:Nj}). Due to the reflection symmetry between the left and right eigenvectors the $k$-dependence in the exponentially decaying prefactors cancels, and the nontrivial part of the overlap comes from $\sin{[(k+1)\varphi_j]}\sin{[(n-k)\varphi_j]}$. Because $\varphi_j=j\pi/n$ those sinuses make $j$ half oscillations as $k$ runs over its values, i.e, the situation is the same as for eigenfunctions of an oscillating string fixed at boundaries. With respect to the reflection around the mid-point eigenfunctions are alternatingly even and odd. This even or odd symmetry of the oscillating part of the eigenvectors (\ref{eq:tLR}) is then responsible for the oscillating sign of $N_j$, and in turn of $c_j$.

The sum of components of $\L_j$ needed in $c_j$ (\ref{eq:cj}) has two parts, one coming from the 1st and the last component of $\L_j$, and the other being equal to the sum of components of eigenvector $\tL_j$ (\ref{eq:tLR}) of matrix $T$. The latter is
\begin{equation}
  \sum_{p=1}^{n-2} [\tL_j]_p=\frac{1-\lph (-1)^j \lambda_j^{n/2-1}}{\lph-\lambda_j} \approx \frac{1}{\lph-\lambda_j},
  \label{eq:Lnorm}
\end{equation}
where one writes $\sin{[(n-k)\varphi_j]}$ as an exponential, resulting in a simple geometrical sum, and where in the approximate sign we neglected $\lambda_j^{n/2-1}$ that is exponentially small in $n$. The first and last components on the other hand give
\begin{equation}
  \frac{\braket{\tL_j}{\mathbf{a}_1+\mathbf{a}_2}}{\lambda_j-1}=\frac{-(-1)^j \lambda_j^{n/2-1}+1}{\lambda_j-1}\approx -\frac{1}{1-\lambda_j},
  \label{eq:L1}
\end{equation}
where again in the approximate sign we dropped exponentially small $\lambda_j^{n/2-1}$. All together this gives
\begin{equation}
  c_j\approx\frac{[\tR_j]_{k-1}}{N_j}\left(\frac{1}{\lph-\lambda_j}-\frac{1}{1-\lambda_j} \right),
  \label{eq:cje}
\end{equation}
and in turn the nonzero eigenvalues contribution to the purity
\begin{widetext}
\begin{equation}
  \Delta I_k(t)=-\frac{2(2\alpha)^{2t+k-n+2}}{n}\sum_{j=1}^{n/2-1}(-1)^j (\cos{\varphi_j})^{2t+k-n+1}\sin{\varphi_j}\sin{(k \varphi_j)} \left( \frac{1}{\lph-\lambda_j}-\frac{1}{1-\lambda_j}\right) = a(t)-b(t),
  \label{eq:dI}
\end{equation}
\end{widetext}
where $a(t)$ is the term with $\frac{1}{\lph-\lambda_j}$ and $b(t)$ the one with $\frac{1}{1-\lambda_j}$. Neglecting $\Aker$ this expression is exact (upto exponentially small terms we dropped), however, it is not very handy for analysis so we are going to rewrite it.

Because $\lph > \lambda_j$ we can expand $\frac{1}{\lph-\lambda_j}=\frac{1}{\lph}\sum_{r=0}^\infty (\lambda_j/\lph)^r$, and interchange the order of the two summations, obtaining
\begin{equation}
  a(t)=\frac{2}{n}\sum_{r=0}^\infty \frac{(2\alpha)^{2t+2r+k-n+2}}{\lph^{r+1}} f_k(2t+2r+k-n+1),
  \label{eq:A}
\end{equation}
where
\begin{equation}
  f_k(p):=-\sum_{j=1}^{n/2-1} (-1)^j \cos^p{\varphi_j}\sin{\varphi_j}\sin{(k \varphi_j)}.
  \label{eq:fk}
\end{equation}
A similar expression with $\lph$ replaced by $1$ is obtain for $b(t)$. The purity can therefore also be written as
\begin{equation}
  \Delta I_k(t)=\frac{4\alpha}{n}\sum_{r=r_{\rm min}}^\infty (2\alpha)^{p}f_k(p)\left(\frac{1}{\lph^{r+1}}-1\right),
  \label{eq:Ifp}
\end{equation}
where $p=2t+2r+k-n+1$ and $r_{\rm min}=0$. This is the main result that we shall repeatedly use.

First, let us look at how it looks like. In Fig.~\ref{fig:I100}(a) the green pluses is Eq.(\ref{eq:Ifp}), and we can see that for $t>\tK$, when the kernel $\Aker$ does not contribute anymore (we shall see that in Section~\ref{sec:kernel}), it gives the exact purity, while for $t \le \tK$ it gives an incorrect very fast exponential decay. One can also observe (data not shown) that the rate of this decay increases with increasing $n$. Let us explain it.

We can see from Eq.(\ref{eq:Ifp}) that for small $t$ and $r$ the power $p$ can be negative, and, because $2\alpha<1$, this gives very large terms that in fact diverge as $n \to \infty$ ($f_k(p)$ is also large for negative $p$, see Appendix~\ref{App:fkp}). Diverging $\Delta I_k(t)$ for $t\le \tK$ is therefore due to terms with non-positive $p$. Those terms are obtained for $r=0,1,\ldots, r_{\rm max}$, with $r_{\rm max}=\lfloor\frac{n-1-k}{2} \rfloor-t=t_{\rm K}-t$ (at $r=r_{\rm max}$ one has $p=-1$ for odd $n-k-1$, and $p=0$ for even $n-k-1$). We shall also rather amusingly see in Sec.~\ref{sec:kernel} that the contribution of the kernel $\Aker^t$ is exactly equal to the negative of the terms with $p\le 0$. Therefore, the kernel part will exactly cancel those diverging terms, so that the full exact expression for the purity (up to exponentially small terms neglected in Eqs.(\ref{eq:Lnorm}-\ref{eq:L1})) is given by Eq.(\ref{eq:Ifp}) taking
\begin{equation}
  r_{\rm min}={\rm max}(0,\tK-t+1),
  \label{eq:rmin}
\end{equation}
instead of $r_{\rm min}=0$! Kernel therefore just renormalizes the sum (\ref{eq:Ifp}) so as to cancel diverging terms.

The dominant diverging term for $t\le \tK$ coming from $A_\lambda$ can be identified from Eq.(\ref{eq:dI}), and is the $j=n/2-1$ term for which $\cos{\varphi_j}$ is the smallest. Collecting the $t$-dependent part we get the leading diverging term in the TDL
\begin{equation}
  \Delta I_k(t) \propto \left(\frac{4\pi\alpha}{n} \right)^{2t}=\left(\frac{4\pi \lph}{(1+\lph)n} \right)^{2t},
  \label{eq:diver}
\end{equation}
where we left out an $n$-dependent prefactor. This is indeed an exponential decay in $t$ (full thin steep black line overlapping with pluses in Fig.~\ref{fig:I100}(a)) but with a rate that increases logarithmically with $n$.

\subsection{Effective decay rate}

Let us now focus on behavior for $t>\tK$ and in particular on an instantaneous effective decay rate $\leff$ defined as $\Delta I(t) \sim \leff^t$, in other words,
\begin{equation}
  \ln{\leff(t)}:= \frac{{\rm d}\ln{\Delta I(t)}}{{\rm d}t}.
  \label{eq:leff}
\end{equation}
We have seen in Fig.~\ref{fig:I100}(b) that $\leff$ is initially equal to $\lph$, while at late times it approaches the 2nd largest eigenvalue $\lambda_1$. In this subsection we are going to show that this is indeed the case, and find the transition time $\tc$ when the phantom decay stops.

First, in the parentheses in Eq.(\ref{eq:Ifp}) one can neglect $1$ compared to $1/\lph^{r+1}$ (because $\lph$ is smaller than $1$, e.g., for $d=2$, when it is the largest, it is $\lph=2/3$, for $d=5$ it is $\lph=5/21\approx 0.24$, and for large $d$ it scales as $\lph \asymp 1/d$), therefore approximating $\Delta I_k(t) \approx a(t)$. Crucial for understanding phantom decay is the function $f_k(p)$. One can show (Appendix~\ref{App:fkp}) that $f_k(p)$ is exactly zero for a bunch of small positive $p$. Specifically, if $k$ is even then $f_k(p)=0$ for odd $p=1,3,\ldots,n-k-3$ (note that for even $k$ the arguments $p$ in Eq.(\ref{eq:Ifp}) are odd), while if $k$ is odd one has $f_k(p)=0$ for even $p=2,4,\ldots,n-k-3$. Because non-positive arguments $p$ are excluded from the summation due to (\ref{eq:rmin}) as they are canceled by the kernel contribution (Section~\ref{sec:kernel}), the first nonzero term in $\Delta I_k(t)$ (\ref{eq:Ifp}) is when $p=n-k-1$. If we denote
\begin{equation}
  \tc=n-k-1,
  \label{eq:tc}
\end{equation}
this nonzero term is obtained for $r=r_1=\tc-t$. If $t<\tc$ then this first $r_1$ is positive, and so the sum in Eq.(\ref{eq:Ifp}) does not start at $r=0$; rather, it starts as $\frac{n}{4\alpha}\Delta I_k(t< \tc)=(2\alpha)^\tc f_k(\tc)/\lph^{\tc-t}+(2\alpha)^{\tc+2} f_k(\tc+2)/\lph^{\tc-t+1}+ \cdots$. Importantly, time dependence does not appear in the argument of $f_k(p)$ but only as a common factor $\lph^t$ in the powers of $\lph$. Therefore one immediately and exactly gets
\begin{equation}
  \leff \approx \frac{a(t+1)}{a(t)}=\lph,\qquad t < \tc.
\end{equation}
We have shown that one has a phantom decay with $\lph>\lambda_1$ for $t<\tc$. The transition time $\tc$ is extensive, i.e., proportional to the system size $n$ as long as $n-k$ is extensive (symmetry between bipartitions $k$ and $n-k$ is broken by the directedness of the staircase configuration, Fig.~\ref{fig:S}). We can also comment on the neglected $1$ in Eq.(\ref{eq:Ifp}) and how much this would cause deviation from $\lph$ -- the neglected relative correction is of order $\lph^{\tc-t}$ and is therefore small for large $\tc-t$, i.e., as long as $t$ is far from the extensive ending point $\tc$ of phantom relaxation.

How about $\leff$ at long times after $\tc$? Let us approximate $a(t)$ only by its first term. While this might not be a very good approximation immediately after $\tc$, we will see that it nevertheless soon gives a good approximation and will furthermore allow us to express the shape of the transition from $\lph$ to $\lambda_1$ in terms of Jacobi theta functions. We therefore approximate
\begin{equation}
  \leff(t) \approx (2\alpha)^2 \frac{f_k(2t+k-n+3)}{f_k(2t+k-n+1)}.
  \label{eq:leffp0}
\end{equation}
From Eq.(\ref{eq:leffp0}) we can also easily get the asymptotic value of $\leff$ for $t \gg \tc$. For very long times one can approximate $f_k(p)$ by its first $j=1$ term in Eq.~(\ref{eq:fk})-- as $t$ grows at fixed $n$ the argument $p$ increases and $\cos^p{\varphi_j}$ get very small, with the largest term being the one with $\varphi_1$. This gives
\begin{equation}
  \leff \longrightarrow (2\alpha)^2 \cos^2{\varphi_1}=\lambda_1.
\end{equation}
That is, we correctly reproduce the asymptotic purity relaxation $\Delta I_k(t) \asymp \lambda_1^t$.

To get the transition shape (Fig.~\ref{fig:I100}(b)) we are going to study
\begin{equation}
\Delta \leff(t) := \leff(t)-\lambda_1.
\end{equation}
As we shall show in next subsections, $\Delta \leff(t)$ has a scaling form for extensive $n-k$,
\begin{equation}
  \Delta \leff(t) \approx \frac{1}{n^2}g(t/n^2),
\end{equation}
with the scaling function $g(x)$ that depends on $k$, but has a short- and long-time forms
\begin{equation}
  g(x \ll 1)\sim \frac{1}{x^2},\qquad g(x) \asymp {\rm e}^{-x}.
\end{equation}
This means that on short times ($t/n \ll n$) the scaling variable is $t/n$, i.e., $\Delta \leff$ will reach any fixed value in a time that scales as $t \sim n$, whereas on long times ($t \sim n^2$) the scaling variable is $t/n^2$. This in particular means that $\Delta \leff(t)$ reaches level $\sim 1/n^2$ on a timescale $t \sim n^2$, explaining a rather slow-looking convergence towards $\lambda_1$ in Fig.~\ref{fig:I100}(b). In the next subsections we are going to calculate the explicit form of the scaling function $g(x)$ for $k=2$, and for $k=n/2$.

\subsection{Transition for bipartition with $k=2$}

\begin{figure}
\centerline{\includegraphics[width=.6\columnwidth]{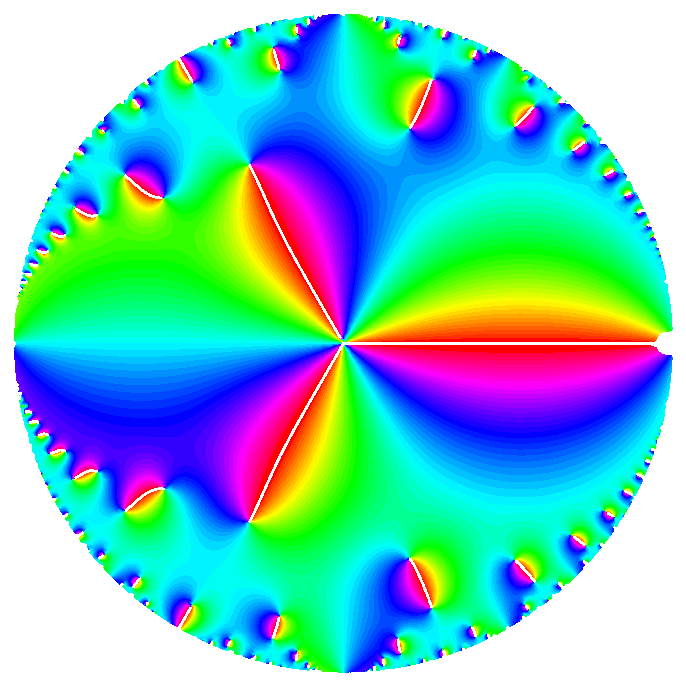}\hskip10pt\includegraphics[width=.1\columnwidth]{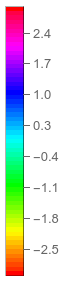}}
\caption{The phase of $q \vartheta''_4(q)/\vartheta'_4(q)$ (\ref{eq:razvojk2}) within the unit circle of complex $|q|<1$. The transition in $\Delta \leff$ for bipartition with $k=2$ is given by this function evaluated at real values of $q \in [0,1]$ (visible as a white horizontal line in figure).}
\label{fig:th4}
\end{figure}

To describe how $\leff$ transitions from $\lph$ to $\lambda_1$ we need to understand $f_2(p)$ (\ref{eq:fk}),
\begin{equation}
  f_2(p)=-2\sum_{j=1}^{n/2-1} (-1)^j \cos^{p+1}{\varphi_j}\sin^2{\varphi_j}.
\end{equation}
We show in Appendix~\ref{App:k2} that one can approximate $f_2(p)$ by the Jacobi theta function~\cite{Abram} $\vartheta_4(z,q)$,
\begin{equation}
  \vartheta_4(z,q)=1+2\sum_{j=1}^\infty (-1)^j q^{j^2}\cos{(2jz)},
  \label{eq:th4}
\end{equation}
or, more precisely, its derivatives with respect to $q$ evaluated at $z=0$,
\begin{eqnarray}
  \vartheta'_4(q)=\frac{{\rm d}\vartheta_4(z=0,q)}{{\rm d}q},\qquad  \vartheta''_4(q)=\frac{{\rm d}\vartheta'_4(q)}{{\rm d}q},
\end{eqnarray}
as
\begin{equation}
  \Delta \leff(t) \approx -(2\alpha)^2 \left(\frac{\pi}{n}\right)^2 c^{2t-n+6} \frac{\vartheta''_4(c^{2t-n+4})}{\vartheta'_4(c^{2t-n+4})},
  \label{eq:razvojk2}
\end{equation}
where we abbreviated $c:=\cos{\varphi_1}=\cos{(\pi/n)}$ and the approximation should be valid in the TDL. The transition from the phantom decay to the asymptotic decay is up-to prefactors given by $q \vartheta''_4(q)/\vartheta'_4(q)$ with real $q=(\cos{\pi/n})^{2t-n+4}$. Theta functions are beautiful objects with an intricate structure in the complex plane as illustrated in Fig.~\ref{fig:th4}.

\begin{figure}
\centerline{\includegraphics[width=.9\columnwidth]{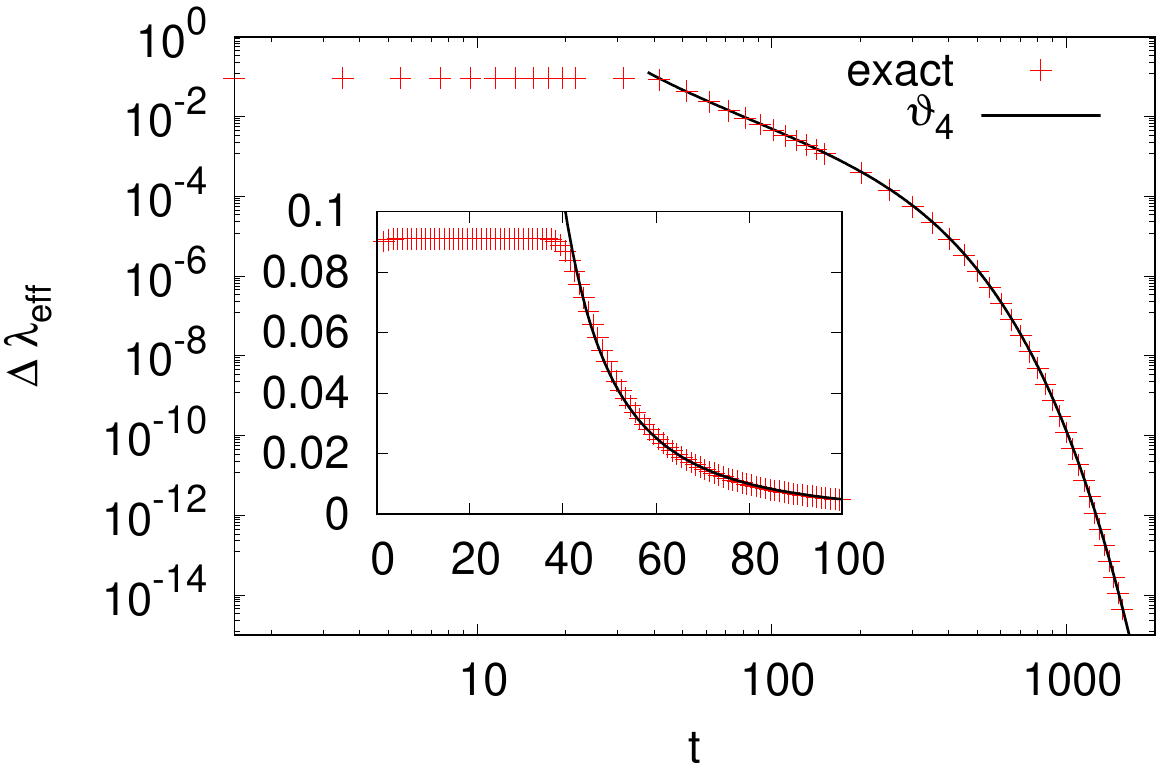}}
\caption{The theoretical decay rate $\Delta \leff$ (thin black curve) given by Eq.(\ref{eq:razvojk2}) agrees well with the exact result (symbols obtained by matrix iteration of $A$) already for $n=40$. Bipartition with $k=2$, and local dimension $d=5$.}
\label{fig:k2}
\end{figure}

In Fig.~\ref{fig:k2} we compare the approximation of $\Delta \leff(t)$ given by Eq.(\ref{eq:razvojk2}) with exact values. We can see that until $\tc$ the effective decay is equal to $\lph$, after which $\Delta \leff$ starts to decrease. Already close to $\tc$ approximation by Eq.~(\ref{eq:razvojk2}) is good even for not very large $n=40$.

\begin{figure}
\centerline{\includegraphics[width=.9\columnwidth]{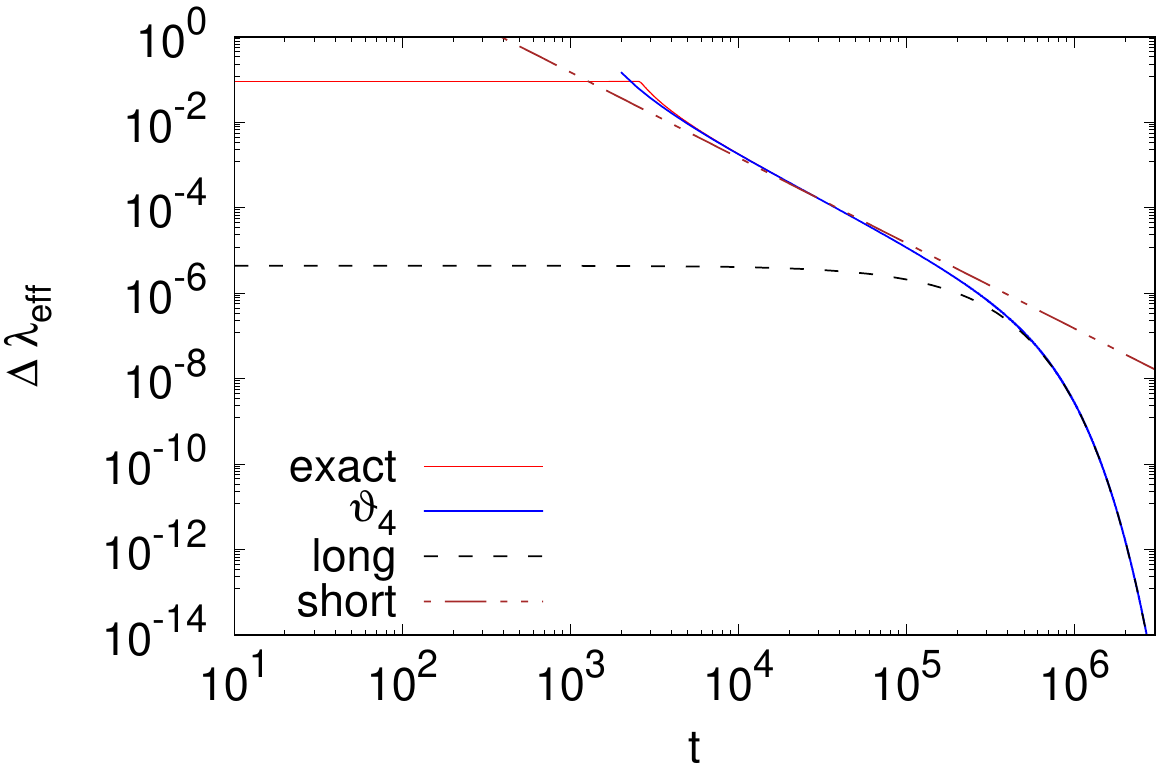}}
\caption{Decay rate $\Delta \leff$ for $n=2000$, $k=2$, and $d=5$. Blue full curve is expression in terms of theta function, Eq.~(\ref{eq:razvojk2}), and we also show the long-time asymptotic exponential form Eq.(\ref{eq:k2long}) (dashed curve), and the short-time power-law of Eq.(\ref{eq:k2short}) (chain curve).}
\label{fig:k2ls}
\end{figure}

One can also simplify Eq.(\ref{eq:razvojk2}) and get its short and long time form. At short times $t \sim n$ we get (Appendix~\ref{App:k2})
\begin{equation}
  \Delta \leff(t) \approx (2\alpha)^2 \frac{1}{4} \left( \frac{n}{t}\right)^2.
  \label{eq:k2short}
\end{equation}
At long times on the scale $t \sim n^2$ one on the other hand gets (Appendix~\ref{App:k2})
\begin{eqnarray}
  \label{eq:k2long}
  \Delta \leff(t) &\asymp& 12(2\alpha)^2 \left(\frac{\pi}{n}\right)^2 \left(\cos{\frac{\pi}{n}}\right)^{6t} \\
  &\approx& 12(2\alpha)^2 \left(\frac{\pi}{n}\right)^2 \exp{\left( -3\pi^2 t/n^2\right)}. \nonumber
\end{eqnarray}
We compare in Fig.~\ref{fig:k2ls} these two expressions, seeing that they agree very well with the exact $\Delta \leff(t)$. Looking more carefully at the convergence with $n$, see Appendix~\ref{App:k2}, one can estimate that the short time power-law approximation (\ref{eq:k2short}), where the scaling variable is $t/n$, holds for $5 \lessapprox t/n \lessapprox 0.02n$, whereas the long time exponential (\ref{eq:k2long}), where the scaling variable is instead $t/n^2$, holds for $t \gtrapprox 0.1n^2$.

\subsection{Transition for bipartition with $k=n/2$}

\begin{figure}
\centerline{\includegraphics[width=.9\columnwidth]{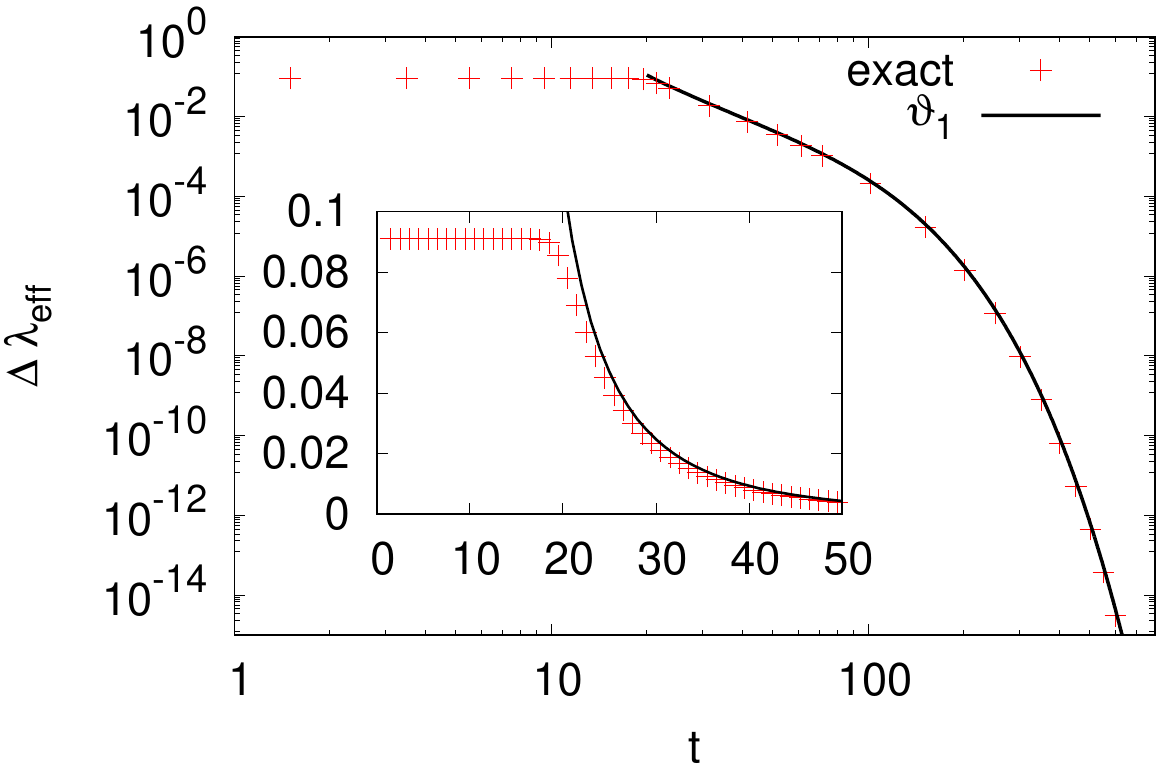}}
\caption{The theoretical decay rate $\Delta \leff$ in Eq.(\ref{eq:razth1}) (full black curve) agrees with the exact result (symbols, obtained by matrix iteration of $A$) already for $n=40$. Bipartition with $k=20$, and $d=5$.}
\label{fig:kn2}
\end{figure}

For half-half bipartition we need $f_{n/2}(p)$ (\ref{eq:fk}), which is
\begin{equation}
  f_{n/2}(p)=-\sum_{j=1}^{n/2-1} (-1)^j \cos^p{\varphi_j}\sin{\varphi_j}\sin{(\frac{\pi}{2}j)}.
  \label{eq:fkn2}
\end{equation}
Using similar techniques as for $k=2$, see Appendix~\ref{App:kn2}, we again get the Jacobi theta function, this time though $\vartheta_1(z,q)$,
\begin{eqnarray}
  \vartheta_1(z,q)&=& 2q^{1/4}\sum_{k=0}^\infty (-1)^k q^{k(k+1)} \sin{[(2k+1)z]}, \nonumber \\
  \vartheta'_1(z,q)&=&{\rm d}\vartheta_1(z,q)/{\rm d}q.
  \label{eq:th1}
\end{eqnarray}
The effective decay is for large $n$
\begin{equation}
  \Delta \leff(t) \approx (2\alpha)^2 \left( \frac{\pi}{n}\right)^2 \left[1-4c^{x_0}\frac{\vartheta'_1(\frac{\pi}{n},c^{x_0})}{\vartheta_1(\frac{\pi}{n},c^{x_0})} \right],
  \label{eq:razth1}
\end{equation}
with $x_0=8t-2n+4$ and $c=\cos{(\pi/n)}$. In Fig.~\ref{fig:kn2} we can see that this approximation agrees with the exact result very well.

Simplifying for short times, one gets (Appendix~\ref{App:kn2})
\begin{equation}
  \Delta \leff(t)\approx (2\alpha)^2 \left(\frac{1}{24}+\frac{1}{8\pi^2} \right)\left( \frac{n}{t}\right)^2,
  \label{eq:kn2short}
\end{equation}
again showing that the scaling variable is $t/n$ with the scaling function being $g(x) \sim 1/x^2$. At long times one instead has
\begin{eqnarray}
  \label{eq:kn2long}
  \Delta \leff(t) &\asymp& 24(2\alpha)^2 \left( \frac{\pi}{n}\right)^2 \left( \cos{\frac{\pi}{n}}\right)^{16t}\\
  &\approx& 24(2\alpha)^2 \left(\frac{\pi}{n}\right)^2 \exp{\left( -8\pi^2 t/n^2\right)}, \nonumber
\end{eqnarray}
and so the scaling variable is $t/n^2$.
\begin{figure}
\centerline{\includegraphics[width=.9\columnwidth]{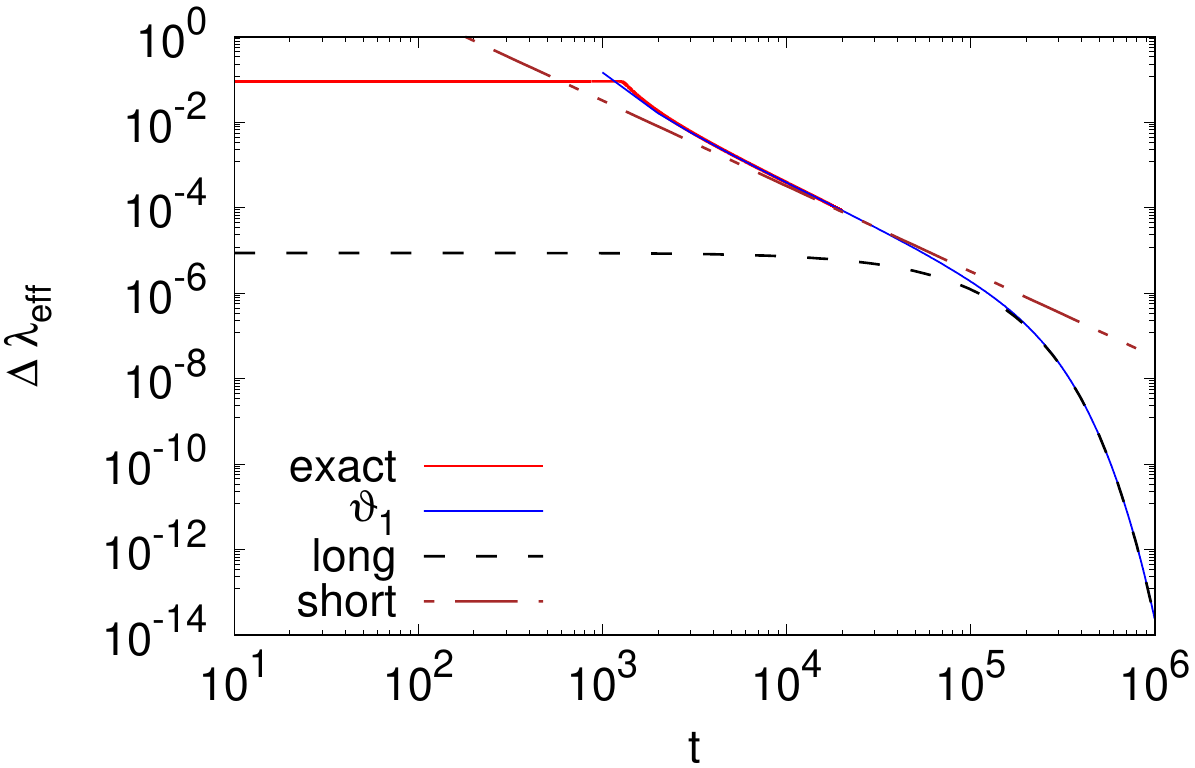}}
\caption{Decay rate $\Delta \leff$ for $n=2000$ and bipartition with $k=1000$, $d=5$. Shown is expression in terms of theta function (blue curve), Eq.~(\ref{eq:razth1}), as well as the long-time asymptotic exponential form Eq.(\ref{eq:kn2long}) (black dashed curve), and the short-time power-law (brown chain line) of Eq.(\ref{eq:kn2short}).}
\label{fig:kn2ls}
\end{figure}
In Fig.~\ref{fig:kn2ls} we see that both approximations agree well with the exact results (no fitting parameters). The short-time expression (\ref{eq:kn2short}) holds in a range $2 \lessapprox t/n \lessapprox 0.01n$, while the long-time expression (\ref{eq:kn2long}) for $t \gtrapprox 0.05n^2$.

\section{Kernel}
\label{sec:kernel}

Here we shall study the kernel part $\Aker$ of $A$ (\ref{eq:Alamker}), specifically the $(n/2-1)$ dimensional kernel of $T$ which will then immediately give us also the $n/2-1$ dimensional kernel of $A$. First, we are going to write down the Jordan normal form of $\Aker$ and study how it contributes to $\Delta I_k(t)$. In the 2nd subsection we are then going to study how a delicate instability of the kernel under small perturbations causes the emergence of the pseudospectrum of $A$, which is one alternative~\cite{PRR} of obtaining the phantom relaxation rate $\lph$.

\subsection{Kernel Jordan normal form}

The kernel is of size $n/2-1$, i.e., half the size of matrix $T$, and has a single Jordan block -- the geometric multiplicity is $1$ and the algebraic $n/2-1$. The spectral decomposition can be written as~\cite{Weintraub}
\begin{equation}
  A_{\rm ker}=\sum_{k=1}^{n/2-2} \ket{\r_k}\bra{\l_{k+1}},
\end{equation}
where we have biorthonormality $\braket{\r_k}{\l_j}=\delta_{k,j}$, $j,k=1,\ldots,n/2-1$, and left and right vectors satisfy the shift property, $A \ket{\r_k}=\ket{\r_{k-1}}$, and $A^{\rm T} \ket{\l_k}=\ket{\l_{k+1}}$. In other words, $\ket{\r_k}$ is in the kernel of $A^p$ for $p\ge k$ (but not for $p<k$), while $\ket{\l_k}$ is in the kernel of $(A^{\rm T})^{p}$ for $p\ge n/2-k$ (but not for $p<n/2-k$). Such vectors are called generalized eigenvectors. If we put $\r_k$ as a $k$-th column of a $n \times (n/2-1)$ rectangular matrix $U$, and $\l_k$ as columns of a $n \times (n/2-1)$ rectangular matrix $V$, biorthonormality becomes $U^{\rm T}V=\mathbbm{1}$, and $V A_{\rm ker} U^{\rm T}=J$, where $J$ is a Jordan matrix of size $(n/2-1)\times (n/2-1)$, i.e. a matrix with $1$ in the first superdiagonal and $0$ everywhere else. 

Due to the shift property the powers of $A_{\rm ker}$ are simple, namely
\begin{equation}
  A_{\rm ker}^t=\sum_{k=1}^{n/2-1-t} \ket{\r_k}\bra{\l_{k+t}},\qquad t \le n/2-2,
  \label{eq:Akert}
\end{equation}
while $A_{\rm ker}^{n/2-1}=0$. Writing $A=A_{\rm ker}+A_{\lambda}$ in terms of the kernel part and the rest, and because $A_{\rm ker}A_{\lambda}=A_{\lambda}A_{\rm ker}=0$, we have $A^t=A_{\rm ker}^t+A_{\lambda}^t$, and we immediately know that the kernel does not play any role for $t\ge n/2-1$ (a specific form of generalized eigenvectors will actually result in even tighter condition).

Denoting by $\ra{k}$ and $\la{k}$, $k=1,\ldots,n/2-1$, the $(n-2)$-component kernel vectors of matrix $T$ (\ref{eq:Ts}),
\begin{equation}
 T_{\rm ker}=\sum_{k=1}^{n/2-2} \ket{\ra{k}}\bra{\la{k+1}},
  \label{eq:Tkerx}
\end{equation}
we can immediately construct the generalized kernel eigenvectors (which have $n$ components) of $A$ as
\begin{equation}
  \r_k=(0,\ra{k},0),\qquad \l_k=(0,\la{k},b_k).
\end{equation}
The scalar $b_k$ satisfies the recursion $b_k=b_{k+1}-\mathbf{a}_2 \cdot \la{k}$, with a starting $b_{n/2}=0$. Dependence on $\alpha$ is rather trivial, namely, by a simple diagonal similarity transformation $T$ can be transformed to a matrix with $\alpha=1$, see Ref.~\cite{PRR}, and so one can get $\ra{k}$ from those for $\alpha=1$ by
\begin{equation}
  [\ra{k}]_j=[\raa{k}]_j\, \alpha^{j-2k},\quad [\la{k}]_j=[\laa{k}]_j\, \alpha^{2k-j},
  \label{eq:ra}
\end{equation}
where we have denoted by $\raa{k}$ and $\laa{k}$ the generalized eigenvectors for $\alpha=1$.

\subsubsection{Right vectors}
\begin{figure}
  \centerline{\includegraphics[width=.8\columnwidth]{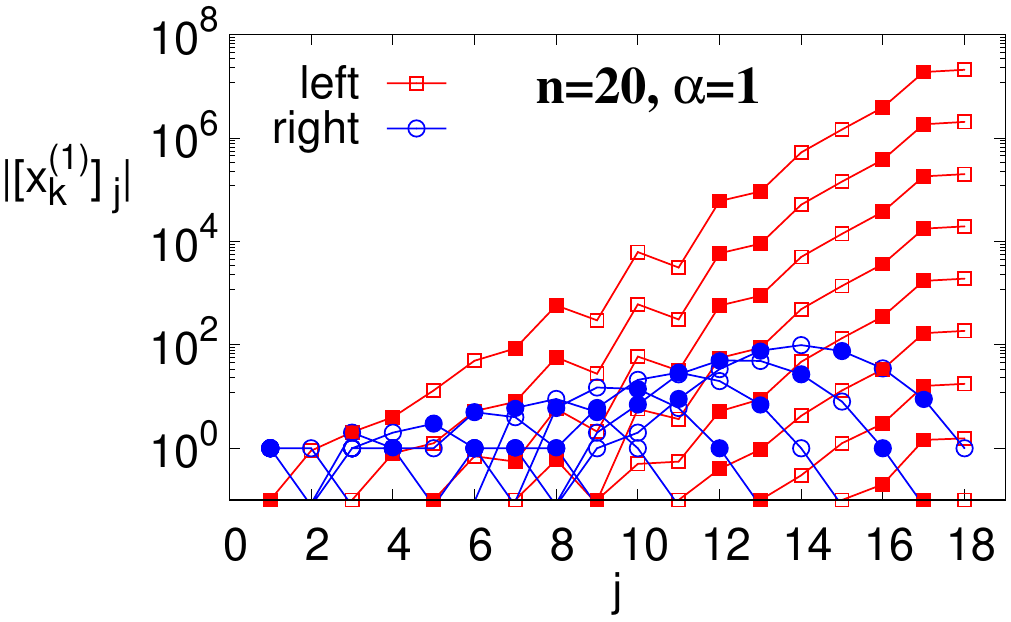}}
\caption{Absolute values of coefficients of left $[\laa{k}]_j$ (Eq.~(\ref{eq:la1}), red squares, $k$ increases from top to bottom) and right $[\raa{k}]_j$ (Eq.~(\ref{eq:ra1}), blue circles, $k$ decreases from right to left) kernel vectors of $T$ for $n=20$, where one has $n/2-1=9$ vectors, each having $n-2=18$ components (full symbols indicate negative values).}
\label{fig:Tkeralpha1}
\end{figure}

Let us now calculate right vectors $\raa{k}$ for $T(\alpha=1)$. Looking at the shift property $T\ket{\raa{k}}=\ket{\raa{k+1}}$ with the specific form of $T$ (\ref{eq:Ts}) we can see that the recursion $[\raa{k}]_j=[\raa{k-1}]_{j-1}-[\raa{k-1}]_{j-2}$ holds. Starting with an appropriate $\raa{1}$ we get
\begin{eqnarray}
  \raa{1} &=&(-1,1,0,\ldots,0)\nonumber \\
  \raa{2} &=&(-1,0,2,-1,0,\ldots,0) \nonumber \\
  \raa{3} &=&(-1,0,1,2,-3,1,0,\ldots,0),\nonumber \\
  \raa{4} &=&(-1,0,1,1,1,-5,4,-1,0,\ldots,0),
\end{eqnarray}
and so on. Recursion can in fact be solved in a closed form, with $\raa{k}$ having non-zero only the first $2k$ components, which are
\begin{equation}
  [\raa{k}]_{2k-p}=-(-1)^{k+p} \sum_{r=0}^{\lfloor p/2 \rfloor} (-1)^r {k-r \choose p-2r},
  \label{eq:ra1}
\end{equation}
where $p=0,\ldots,2k-1$ and $\lfloor x \rfloor$ is the largest integer that is smaller or equal than $x$. We for instance see that the last nonzero component is always $[\raa{k}]_{2k}=-(-1)^k$, the 2nd last is $[\raa{k}]_{2k-1}=(-1)^k k$, the 3rd last is quadratic in $k$, etc.. The modulus of $[\raa{k}]_j$ increases with decreasing $j$, with the maximum reached around index $j\approx 2k-k/2$. At large $k$ vectors $[\raa{k}]_j$ are therefore peaked close to the right edge (Fig.~\ref{fig:Tkeralpha1}), however, including the dependence on $\alpha$ (\ref{eq:ra}) will cause an exponential decay with $j$ and therefore $\ra{k}$ themselves are localized at the left edge (Fig.~\ref{fig:Tkeralpha}).
Because $[\ra{p}]_{k}$ is nonzero only for $k\le 2p$, and $[\r_p]_k=[\ra{p}]_{k-1}$, we see that we will have a non-zero contribution to the purity $I_k(t)$, Eq.(\ref{eq:Iker}), only if $k-1 \le 2(\frac{n}{2}-t-1)$, i.e.,
\begin{equation}
  t \le \tK=\left\lfloor \frac{n-k-1}{2} \right\rfloor.
  \label{eq:tKK}
\end{equation}
For bipartition $k$ the kernel does not contribute for $t> \tK$.

\subsubsection{Left vectors}

\begin{figure}
  \centerline{\includegraphics[width=.8\columnwidth]{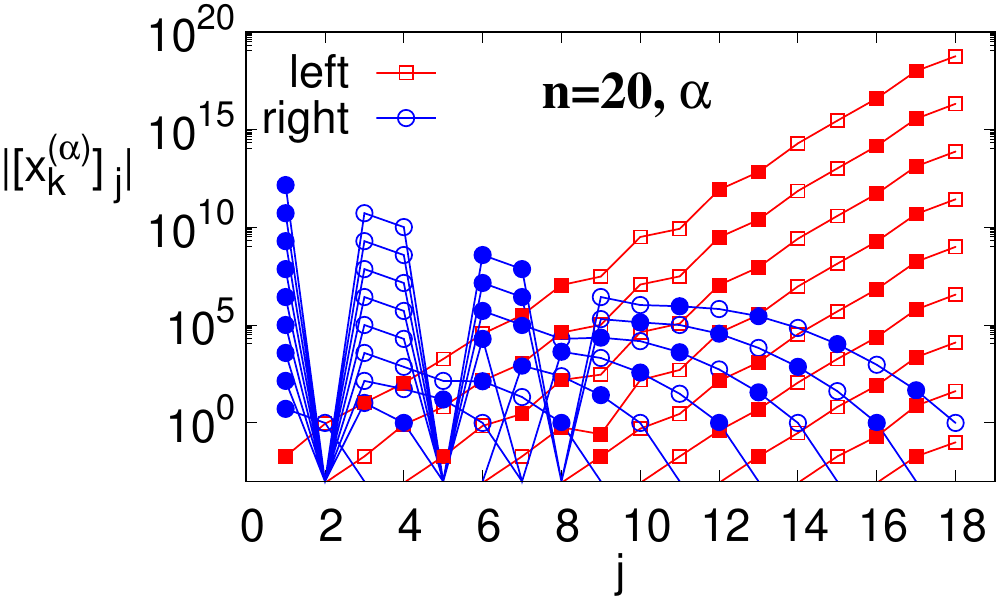}}
\caption{The same vectors as in Fig.~\ref{fig:Tkeralpha1}, this time for nonzero $\alpha=5/26$ ($d=5$, $n=20$), showing the non-Hermitian skin effect of Jordan normal form kernel generalized eigenvectors (\ref{eq:Tkerx}). Left are localized at the right edge (red squares), right (blue circles) at the left edge (for squares $k$ increases from top to bottom, for circles $k$ increases from bottom to top).}
\label{fig:Tkeralpha}
\end{figure}
We have seen that all $\r_{k}$ are $n$-independent with a compact explicit form. The left vectors $\l_k$ are on the other hand more complicated. Because $T^{\rm T} \ket{\laa{n/2-1}}=0$ the recursion starts with the last index $k=n/2-1$, for which we have $\laa{n/2-1}=(0,\ldots,0,-\frac{2}{n},\frac{2}{n})$, i.e., nonzero are only the last two components whose value $\pm \frac{2}{n}$ is fixed by normalization $\braket{\raa{n/2-1}}{\laa{n/2-1}}=1$. This normalization is what brings in the $n$-dependence (due to the $n$-dependent 2nd to last component $[\raa{n/2-1}]_{n-3}= \pm (n/2-1)$), which, as we will see, then trickles into an increasingly more complicated $n$ dependence of subsequent left vectors $\laa{n/2-k}$. Due to the structure of $T^{\rm T}$ they have non-zero values only in the last $2k$ components, and they satisfy the recursion $[\laa{k-1}]_j=[\laa{k}]_{j+1}-[\laa{k}]_{j+2}$. This recursion is enough the get all components of $\laa{k-1}$ apart from the last two ones, say $a$ and $b$, which must instead be determined from the biorthogonality $\braket{\raa{k}}{\laa{k-1}}=0$, and $T^{\rm T}\ket{\laa{k-1}}=\ket{\laa{k}}$ which for the last component results in $a+b=[\laa{k}]_{n-2}$. This for instance gives the last two vectors,
\begin{eqnarray}
    \label{eq:la1}
  \laa{n/2-1}&=&(-1)^{n/2}(0,\ldots,0,-\frac{2}{n},\frac{2}{n}) \\
  \laa{n/2-2}&=&(-1)^{n/2}(0,\ldots,0,\frac{2}{n},-\frac{4}{n},-\frac{n^2-52}{12n},\frac{n^2-28}{12n}). \nonumber
\end{eqnarray}
Polynomials in the numerators of the last two components of $\laa{n/2-k}$ are of order $k$ and get more complicated for increasing $k$. We were not able to obtain any nice closed form for them, so we just show them in Fig.~\ref{fig:Tkeralpha1}. We can see that they are all localized at the right edge, similarly as $\alpha$-dependent vectors $\l_k$ (Fig.~\ref{fig:Tkeralpha}). 

Once we have left and right kernel vectors we can use $\Aker^t$ (\ref{eq:Akert}) to write down the contribution $I_k^{\rm ker}$ to the purity coming from $\Aker$. Using a product initial state $\bI(0)$ (\ref{eq:I0}) and Eq.(\ref{eq:Ik}) we have 
\begin{equation}
  I_k^{(\rm ker)}(t)= \sum_{p=1}^{n/2-t-1}[\r_p]_k \sum_{j=1}^{n}[\l_{p+t}]_j.
  \label{eq:Iker}
\end{equation}
Observe that, as opposed to $A_\lambda$, the time dependence comes in here only implicitly via the number of terms we have in the sum, and in which left vector is multiplied with which right vector. Nevertheless, as we will see, because of extensive size of the kernel, time dependence of $I_k^{({\rm ker})}(t)$ is rather interesting. The sum of components of $\l_p$ become increasingly more complicated for decreasing $p$, see Eq.(\ref{eq:la1}). The first two are for instance $\braket{\l_{n/2-1}}{\bI(0)}=(-1)^{n/2}(\frac{2}{n}-\frac{4\alpha}{n})$, and $\braket{\l_{n/2-2}}{\bI(0)}=(-1)^{n/2}(-\frac{4}{n}-\frac{n^2-52}{12n\alpha}+\frac{n^2-28}{12n\alpha^2}(1-\alpha))$. Using this two explicit forms we can write down an exact contribution for the two largest times when the kernel still contributes, i.e., $I_k^{(\rm ker)}(\frac{n}{2}-2)=[\r_1]_k \braket{\l_{n/2-1}}{\bI(0)}$, and $I_k^{(\rm ker)}(\frac{n}{2}-3)=[\r_1]_k \braket{\l_{n/2-2}}{\bI(0)}+[\r_2]_k \braket{\l_{n/2-1}}{\bI(0)}$. At time $n/2-2$ only bipartitions with $k=2,3$ get the kernel contribution (\ref{eq:tKK}), while at $t=n/2-3$ bipartitions $k=2,\ldots,5$ do. We do not write all expressions out but to nevertheless give a flavor we get for $t=n/2-2$
\begin{eqnarray}
  \label{eq:Ilast}
  I_2^{(\rm ker)}(\frac{n}{2}-2)&=& \frac{2}{n}(-1)^{n/2}\left( 1-\frac{1}{\lph}\right) \\
  I_3^{(\rm ker)}(\frac{n}{2}-2)&=&-\frac{2\alpha}{n}(-1)^{n/2} \left( 1-\frac{1}{\lph}\right).
  \label{eq:I2ndlast}
\end{eqnarray}
Expressions are relatively simple and similar looking to the spectral contribution in Eq.(\ref{eq:Ifp}). In fact, evaluating terms in Eq.(\ref{eq:Ifp}) for non-positive argument $p=-1$ -- for $t=n/2-2$ there is only one term in the sum having $r=0$ -- we get $\Delta I_2(n/2-2)=\frac{4\alpha}{n}(2\alpha)^{-1} f_2(-1)(1/\lph-1)$. Using the value of $f_2(-1)$ (\ref{eq:f2}) one gets exactly $-I_2^{(\rm ker)}(\frac{n}{2}-2)$ (\ref{eq:Ilast}). A diverging term coming from the spectrum of $A_\lambda$ for a non-positive argument $p=-1$ exactly cancels the kernel contribution. Similar cancellation is observed also for $\Delta I_3(n/2-2)$, as well as for $I_{k=2,3,4,5}(n/2-3)$. We have also checked symbolically some other values of $t$ and $k$, as well as numerically all possible $k$ and $t$ for a range of system sizes, and we always find an exact cancellation. We conjecture that the kernel contribution to the purity (\ref{eq:Akert}) is due to the specific localized form of left and right generalized eigenvectors always exactly equal to the negative of the spectral contribution from $A_\lambda$ (\ref{eq:Ifp}) with non-positive $p$. Those terms are obtained for $r$ in Eq.(\ref{eq:Ifp}) in the range
\begin{equation}
  r=0,1,\ldots,r_{\rm max}=\tK-t.
\end{equation}
In other words, the kernel contribution is automatically accounted for if we take the summation in Eq.(\ref{eq:Ifp}) not from $r=0$, but rather from $r_{\rm min}=r_{\rm max}+1=\tK-t+1$ (\ref{eq:rmin}) such that arguments of $f_k(p)$ are positive.

We have seen that without taking into account the kernel, we do not get the correct decay for $t<\tK$ (green pluses in Fig.~\ref{fig:I100}(a)). To correctly describe the phantom decay at those times one needs the kernel. One can say that the phantom decay $\lph$ is somehow encoded in the structure of localized vectors of kernel Jordan normal form. At the same time we have also seen that $\lph$ is also encoded in the nonzero spectrum part -- for $t>\tK$ we can get $\lph$ also out of $A_\lambda$ alone. Simple-looking matrix $A$, or equivalently $T$ (\ref{eq:Ts}), whose spectrum does not contain any signs of $\lph$ that governs the correct asymptotic relaxation rate, has $\lph$ encoded in the structure of $n/2-1$ eigenvectors corresponding to non-zero eigenvalues, or, independently, also in the structure of the Jordan normal form kernel of size $n/2-1$. In short, it is not (just) the eigenvalues but rather eigenvectors that are crucial.

\subsection{Kernel breakdown and the pseudospectrum}

In the previous section we have seen how $\Aker$ and $A_\lambda$ are both crucial to correctly describe phantom decay for $t<\tK$, while at longer times only $A_\lambda$ is required to recover $\lph$. In both regimes the eigenvectors (and not just the eigenvalues) are the ones that determine the relaxation rate.

In Ref.~\cite{PRR} it has been observed that the phantom decay $\lph$ is exactly equal to the norm of the pseudospectrum. One definition of a pseudospectrum~\cite{trefethen}, more precisely $\varepsilon$-pseudospectrum, of a matrix $T$ is that it is equal to the spectrum of a perturbed matrix (supremum of spectra for all $\| E \| \le 1$)
\begin{equation}
  T(\varepsilon)=T+\varepsilon\cdot E.
  \label{eq:Teps}
\end{equation}
Fixing small $\varepsilon$ the pseudospectrum is well defined and independent of $\varepsilon$ in the TDL. For normal matrices the pseudospectrum is equal to the spectrum, however, for non-Hermitian ones the two can be different. How does the pseudospectrum, and therefore $\lph$, emerge from the spectrum of $A$ under small perturbation?

For a start let us remind ourselves of the standard perturbation theory~\cite{Wilkinson}. A non-degenerate eigenvalue $\lambda_j$ and eigenvector change as
\begin{eqnarray}
  \lambda_j(\varepsilon)&=&\lambda_j+\varepsilon \cdot \bracket{\tL_j}{E}{\tR_j}+\cdots,\\
  \ket{\tR_j(\varepsilon)} &=& \ket{\tR_j}+\varepsilon \sum_{k\neq j} \frac{\bracket{\tL_k}{E}{\tR_j}}{\lambda_j-\lambda_k}\ket{\tR_k}+\cdots
  \label{eq:pert1}
\end{eqnarray}
One can bound the eigenvector difference by (assuming $||E||=1$)
\begin{equation}
  \| \ket{\tR_j(\varepsilon)}-\ket{\tR_j} \| \le \varepsilon \cdot \sum_{k \neq j}\frac{\| \tL_k \|\cdot \|\tR_j\|}{|\lambda_j-\lambda_k|} \| \tR_k \| .
\end{equation}
For Hermitian matrices left and right eigenvectors are the same and normalized and therefore a large change can be achieved only if some other eigenvalue is close to $\lambda_j$ -- that is why we often focus on spectral properties like gaps. However, we can see that for non-Hermitian matrices there is another option, namely, the norms of eigenvectors can diverge and this is exactly what is happening for our $T$.

\begin{figure}[t]
  \centerline{\includegraphics[width=.8\columnwidth]{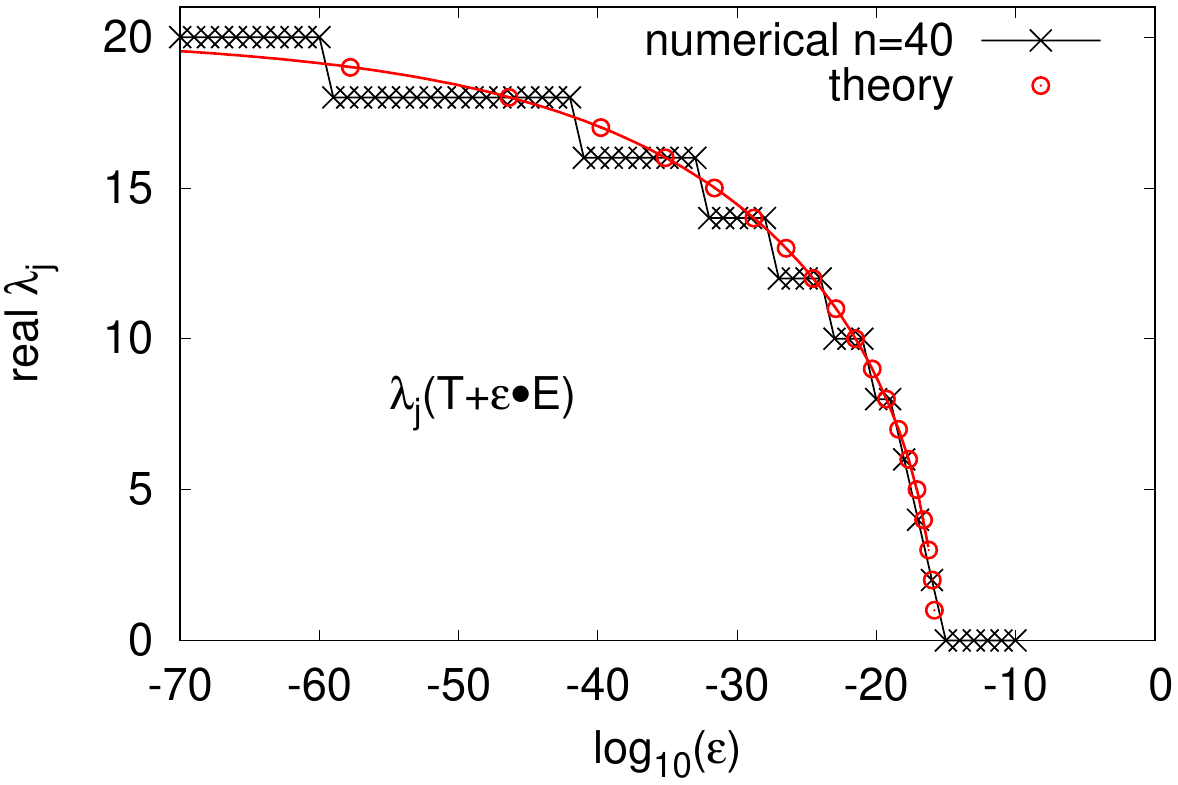}}
\caption{Number of real eigenvalues of $T+\varepsilon E$ for a single realization of $E$, and $n=40, d=5$. Red curve (circles) is theory given by Eq.(\ref{eq:pertth}) without any fitting parameters.}
\label{fig:kernel40}
\end{figure}
\begin{figure*}
  %  \centerline{\includegraphics[width=.6\columnwidth]{kernel_spekter_n20.pdf}\includegraphics[width=1.4\columnwidth]{kernel_combined.pdf}}
  \centerline{\includegraphics[width=.42\columnwidth]{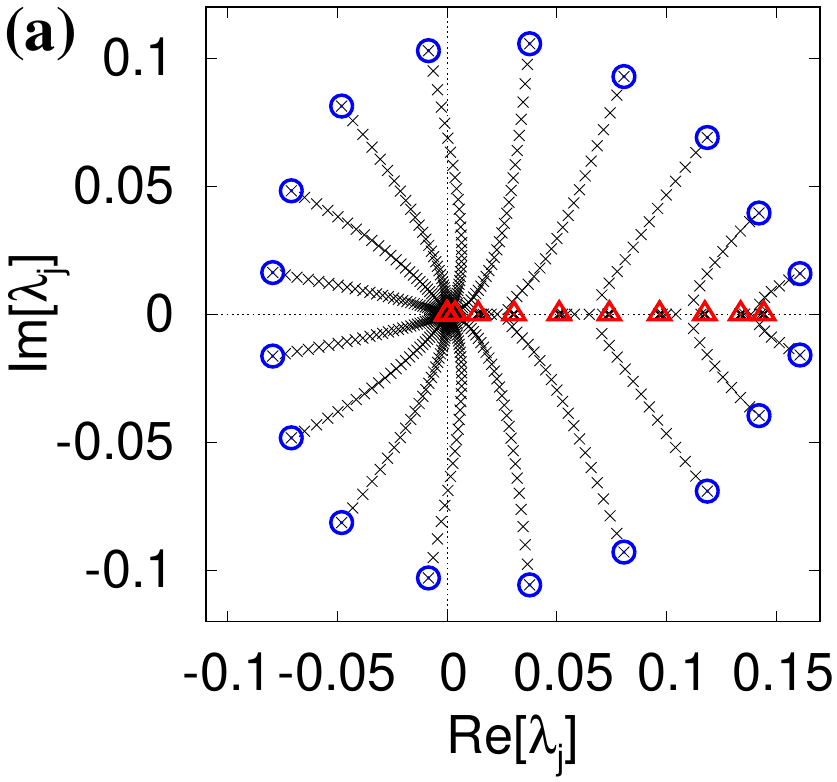}\includegraphics[width=0.53\columnwidth]{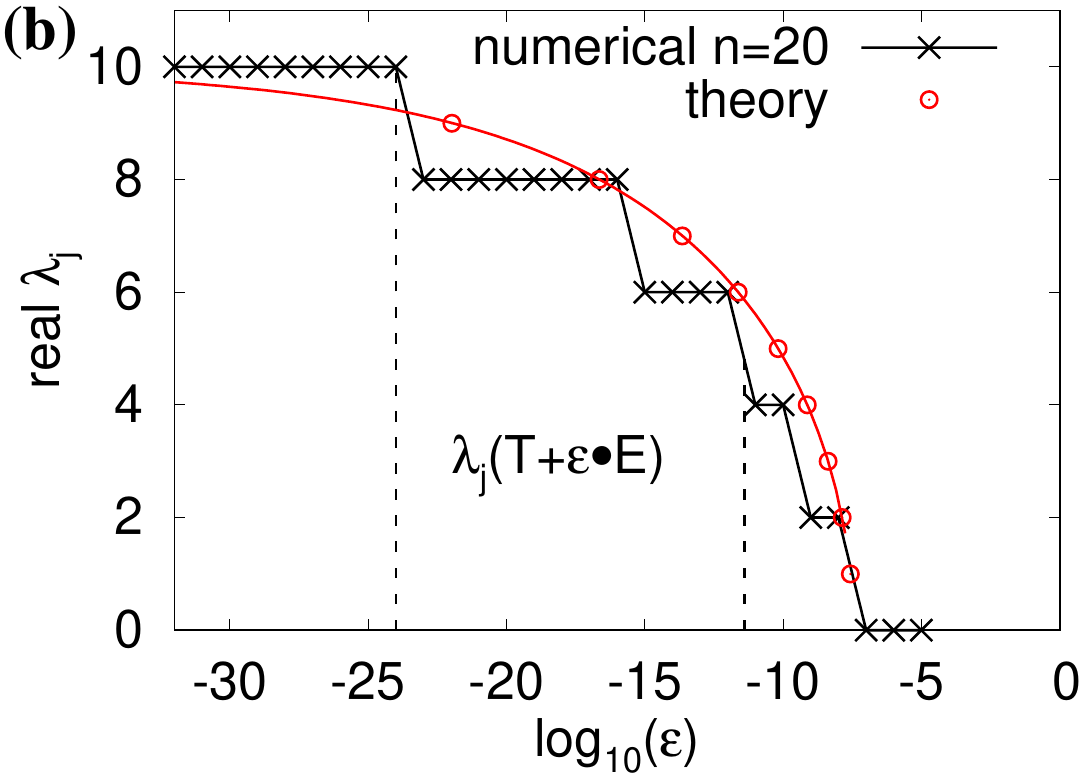}\includegraphics[width=0.65\columnwidth]{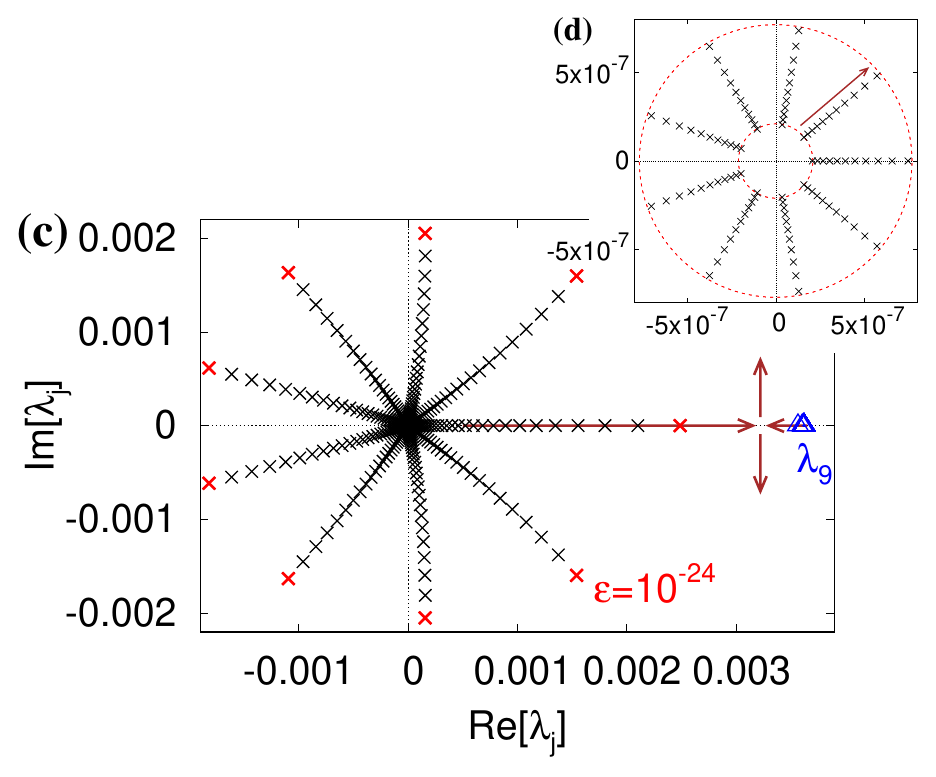}\includegraphics[width=0.55\columnwidth]{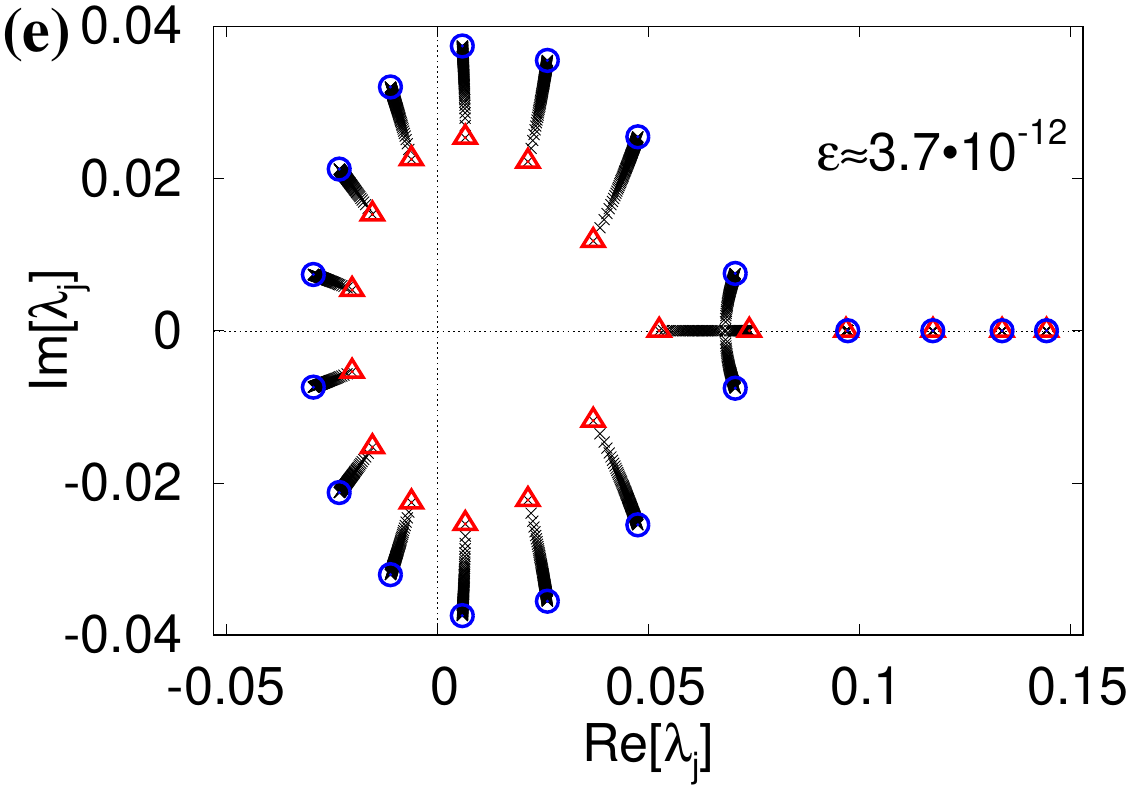}}
%  \centerline{\includegraphics[width=.8\columnwidth]{kernel_n20.pdf}}
%  \centerline{\includegraphics[width=.8\columnwidth]{kernel_spekter_n20.pdf}}
%  \centerline{\includegraphics[width=.8\columnwidth]{kernel_explosion_n20.pdf}}
%  \centerline{\includegraphics[width=.8\columnwidth]{kernel_zoomexplosion_n20.pdf}}
%  \centerline{\includegraphics[width=.8\columnwidth]{kernel_collision_n20.pdf}}
\caption{Spectrum evolution of $T+\varepsilon\cdot E$ (single instance of $E$, $n=20, d=5$). (a) Spectrum for $\varepsilon=10^{-x}$ and $x=5,5.5,6,\ldots,60$ (black crosses, with the largest $\varepsilon=10^{-5}$ with blue circles, and smallest $\varepsilon=10^{-60}$ with red triangles). (b) Similar to Fig.~\ref{fig:kernel40} but with $n=20$. Vertical dashed lines mark perturbation $\varepsilon=10^{-24}$ when the first collision shown in (c) happens, and $\varepsilon=3.7\cdot 10^{-12}$ of frame (e). (c) Exploding kernel eigenvalues for small $\varepsilon$ until just before the first collision with the smallest non-zero $\lambda_9=(2\alpha \sin\frac{\pi}{n})^2$ happens (brown arrows). (d) Zoom-in on $\varepsilon=10^{-x}$ for $x=55,\ldots,60$, two red circles are $\varepsilon^{1/9}$ for the smallest and the largest shown perturbation (\ref{eq:pert}). (e) Collision of two eigenvalues around $\varepsilon \approx 3.7\cdot 10^{-12}$. Shown are spectra for $\varepsilon=y\cdot 10^{-13}$ with $y=1,2,\ldots,100$ (dense black crosses, from red triangles for $y=1$, to blue circles for $y=100$). The remaining 4 real eigenvalues barely move in this range of $\varepsilon$.}
\label{fig:kernel_spec}
\end{figure*}

However, there is another twist in which the kernel, due to its extensive size $n/2-1$, plays a crucial role. Namely, while the standard perturbation theory (\ref{eq:pert1}) predicts that the effect of perturbation is linear in $\varepsilon$, albeit can be large due to localized eigenvectors, a degenerate perturbation theory says that a degenerate eigenvalue $\tilde{\lambda}_j$ of algebraic multiplicity $p$ behaves under perturbation in a non-linear way (Poiseux series)~\cite{trefethen,Kato},
\begin{equation}
  \tilde{\lambda}_j(\varepsilon)=\tilde{\lambda}_j+ b\,\varepsilon^{1/p}\, {\rm e}^{\ii 2\pi j/p} +\cdots.
  \label{eq:deg}
\end{equation}
In our case the degenerate eigenvalue in question is the zero eigenvalue of the kernel, $\tilde{\lambda}_j=0$ with $j=1,\ldots,p$ and $p=n/2-1$. Because of this nonlinearity shifts of zero eigenvalues scale as $\varepsilon^{1/(n/2-1)}$ and will be for large $n$ much larger than the shift of non-degenerate non-zero eigenvalues (\ref{eq:pert1}). Already under very small perturbation the kernel will rapidly explode in $n/2-1$ rotationally symmetric rays (\ref{eq:deg}). The expanding front of eigenvalues will then hit the non-zero $\lambda_j$, causing an exceptional point~\cite{Ueda20}, after which the real eigenvalues are ejected into a complex plane, eventually resulting in the pseudospectrum. Namely, for our matrix the pseudospectrum is an oval-shaped curve in the complex plane, e.g. seen as blue points in Fig.~\ref{fig:kernel_spec}(a), or in Ref.~\cite{PRR}. It is therefore the nonlinear instability of the kernel under perturbation (\ref{eq:deg}) that will govern the spectrum of $T(\varepsilon)$. 

We can estimate the perturbation strength $\varepsilon$ when the $\lambda_j$ will be ejected from the real axis by
\begin{equation}
  \lambda_j=\left(2\alpha \cos{\frac{\pi j}{n}}\right)^2=\varepsilon^{1/(n/2-1)}.
  \label{eq:pert}
\end{equation}
The pseudospectrum is therefore expected to be established roughly when
\begin{equation}
  \varepsilon \sim (2\alpha)^{n-2},
\end{equation}
which is exponentially small in $n$. In Fig.~\ref{fig:kernel40} we compare numerically computed~\cite{foot3} number of real eigenvalues at different $\varepsilon$ for a single realization of $T(\varepsilon)$ (\ref{eq:Teps}), with $E$ being a matrix of real Gaussian numbers of zero mean and unit variance, with theory from (\ref{eq:pert}), i.e.,
\begin{equation}
  j=\frac{n}{\pi} \arccos{\left(\frac{\varepsilon^{1/(n-2)}}{2\alpha} \right)},
  \label{eq:pertth}
\end{equation}
that predicts the number of real eigenvalues as a function of perturbation strength $\varepsilon$. We can observe very good agreement.

In Fig.~\ref{fig:kernel_spec} we illustrate the emergence of the pseudospectrum, that is, follow the changing spectrum of $T(\varepsilon)$ as $\varepsilon$ varies, again for a single realization of a Gaussian perturbation matrix $E$. We can see that the scenario of expanding kernel eigenvalues described by Eq.(\ref{eq:deg}), taking simply $b=1$, describes collisions between real eigenvalues (\ref{eq:pert}) rather well. 

\section{Conclusion}

We studied the average purity evolution in a random staircase circuit with open boundary conditions. The process can be described by an iteration of a non-symmetric matrix whose size is equal to the number of qudits $n$. Contrary to intuition, the relaxation rate in the thermodynamic limit is not given by the finite gap, but rather is slower, that is, slower than any non-steady-state eigenvalue would suggest. Using an exact spectral decomposition we explain how such phantom relaxation rate emerges (i) out of a non-degenerate spectrum and in particular due to localized eigenvectors (non-Hermitian skin effect), and (ii) how extensively large Jordan normal form of the kernel, where the generalized eigenvectors are also localized, likewise results in phantom relaxation at short times. In both cases the non-hermitian skin effect manifests itself in exponentially large expansion coefficients that alternate in signs, almost mutually canceling, and together resulting in a slow decay.

This shows that the time-dependence under non-Hermitian matrices can, up-to times extensive in $n$, result in a behavior that would not be possible under Hermitian dynamics. Such behavior can be caused either by a completely non-degenerate ordinarily looking spectrum, or, alternatively, also by a large non-diagonalizable Jordan block. In both cases it is due to properties of (generalized) eigenvectors. In a non-Hermitian system it is the eigenvectors and a delicate cancellation and not eigenvalues that are important.

A similar conclusion is reached also from another viewpoint. Namely, it is well known in mathematics~\cite{trefethen} that the spectrum of a non-Hermitian matrix can be very sensitive to small perturbations. Because this sensitivity can be exponentially large in $n$, it is the pseudospectrum rather than the spectrum that one needs to look at. Because of an extensively large Jordan normal form the spectrum depends non-linearly on the perturbation strength $\varepsilon$ as $\varepsilon^{1/(n/2-1)}$, so that the spectrum changes to the pseudospectrum already at an exponentially small perturbation strength.

It would be interesting to study such phantom relaxation in higher dimensions, or in the presence of longer-range couplings where entanglement dynamics can be different. In additon, we expect similar behavior also for quantities different than purity. For instance, OTOC relaxation under the brickwall random circuit with periodic boundary conditions results in a non-symmetric tridiagonal matrix~\cite{tobe} having similar underlying mathematics. While our exact solution pertains to the specific purity setting and relies on the simplicity of the resulting Toeplitz matrix (\ref{eq:Ts}), it should be of use in other non-Hermitian situations. The non-Hermitian skin effect arises because of the asymmetry in $T$ (\ref{eq:Ts}) coming in turn from the asymmetry in relaxation between different bipartitions $k$ in $I_k$: purities with smaller $k'<k$ feed into $I_k$ with the amplitude decaying with $k-k'$ (see also Fig.~\ref{fig:I100}(b)). There is an emergent directionality in the relaxation process (i.e., an ``arrow of time''). An interesting question is if the specific behavior identified here is more generic in relaxation where one also expects an effective breaking of time reversibility.

\begin{acknowledgements}
I acknowledge support by Grants No.~J1-4385 and No.~P1-0402 from the Slovenian Research Agency.
\end{acknowledgements}

%---------------------------
\clearpage
\appendix

\section{Properties of $f_k(p)$}
\label{App:fkp}

We would like to show that (assuming even $n$, positive integers $k$ and $r$, and $\varphi_j=\frac{\pi j}{n}$)
\begin{eqnarray}
  f_k(p)&=&-\sum_{j=1}^{n/2-1} (-1)^j \cos^p{\varphi_j}\sin{\varphi_j}\sin{(k \varphi_j)},\nonumber \\
  \label{eq:ksod}
  f_{2r}(2p-1)&=&0 \quad \hbox{for }p=1,2,\ldots,\frac{n}{2}-r-1 \\
  f_{2r-1}(2p)&=&0 \quad \hbox{for }p=1,2,\ldots,\frac{n}{2}-r-1.
  \label{eq:klih}
\end{eqnarray}
First, we rewrite a product of two sinuses in terms of cosines, obtaining
\begin{eqnarray}
  \label{eq:hk}
  2f_k(p)&=&h_{k+1}(p)-h_{k-1}(p),\\
  h_r(p)&=&\sum_{j=1}^{n/2-1} (-1)^j \cos^p{\left(\frac{\pi j}{n}\right)}\cos{\left(\frac{r \pi j}{n}\right)}. \nonumber
\end{eqnarray}
The goal is to show that $h_k(p)$ is constant in an appropriate range of $p$ and $r$, directly implying Eqs.(\ref{eq:ksod}) and (\ref{eq:klih}).

{\em Even bipartition $k$.--} Let us consider even $k=2r$, for which we need $h_{2r\pm 1}(p)$ with odd $p$, that is $h_{2r-1}(2p-1)=\sum_{j=1}^{n/2-1} (-1)^j b_j(n)$, with
\begin{equation}
   b_j(n)=\cos^{2p-1}{\left(\frac{\pi j}{n} \right)}\cos{\left( \frac{(2r-1)\pi j}{n}\right)}.
\end{equation}
One has $b_{n/2}(n)=0$ and $b_{n-j}(n)=b_j(n)$, and therefore we can extend the sum over $j$ to $n-1$, getting
% b_n=1, b_0=1
\begin{equation}
  2h_{2r-1}(2p-1)=\sum_{j=1}^{n-1} (-1)^j b_j(n).
\end{equation}
Separately summing all even $j$, and all odd $j$, we can write
\begin{eqnarray}
  h_{2r-1}(2p-1)&=&B_{\frac{n}{2}}-\frac{1}{2}B_n,\\
 B_m&=&\sum_{j=1}^{m-1}\cos^{2p-1}{\left(\frac{\pi j}{m} \right)}\cos{\left( \frac{(2r-1)\pi j}{m}\right)}. \nonumber
\end{eqnarray}
Let us consider $1+B_m=b_0(m)+B_m$. The benefit of that is that the sum over $j=0,\ldots,m-1$ of $\exp{(\ii \frac{\pi}{m} j 2p)}$ is zero for all integer $p$ except for integer multiples of $m$,
\begin{eqnarray}
  w(p)&=&\sum_{j=0}^{m-1} {\rm e}^{\ii \frac{\pi}{m} j 2p}={\rm e}^{\ii p \frac{m-1}{m}}\frac{\sin{(p \pi)}}{\sin{\frac{p \pi}{m}}},\nonumber \\
  w(p \in \mathbb{Z})&=&
  \begin{cases}
    m & p=ml, l\in \mathbb{Z}\\
    0 & \hbox{else}
    \end{cases}.
\end{eqnarray}
Taking $b_j(m)$ and expressing cosines in terms of an exponential, and using binomial expansion for the $(2p-1)$-th power, we have
\begin{equation}
1+B_m=\frac{1}{2^{2p}}\sum_{t=0}^{2p-1} {2p-1 \choose t}\{w(t+r-p)+w(t-r-p+1)\}.
\end{equation}
Because both arguments of $w$ are integers, we will have a nonzero term only if one of the two arguments is a multiple of $m$. For positive $r$ and $p<m-r$ the two arguments can be only $0$, but not e.g. $-m$. This means that either $t=p-r$, or $t=r+p-1$. Taking into account that one needs also to have $2p-1 \ge t\ge 0$, which gives $p\ge r$, finally results in
\begin{equation}
  1+B_m=
  \begin{cases}
    0 &,\hbox{for }p=1,\ldots,r-1,\\
    \frac{m}{2^{2p-1}} {2p-1 \choose p-r}&,\hbox{for }p=r,\ldots,m-r.
  \end{cases}
\end{equation}
For larger $p$ one would have a sum of more binomial symbols. The precise form of $B_m$ is actually not important, what matters is that for $p=1,2,\ldots,n-r$ one has $B_m \propto m$, from which one immediately gets
\begin{equation}
  h_{2r-1}(2p-1)=-\frac{1}{2},\quad \hbox{for }p=1,2,\ldots,\frac{n}{2}-r.
\end{equation}
Using Eq.(\ref{eq:hk}) this proves Eq.(\ref{eq:ksod}).

{\em Odd bipartition $k$.--} Writing $k=2r-1$ we need $h_{2r}(2p)$ in Eq.(\ref{eq:hk}),
\begin{equation}
  h_{2r}(2p)=\sum_{j=1}^{n/2-1} (-1)^j \cos^{2p}{\left(\frac{\pi j}{n} \right)}\cos{\left( \frac{2r\pi j}{n}\right)}.
\end{equation}
By a similar procedure as before we get
\begin{eqnarray}
  h_{2r}(2p)&=&C_{\frac{n}{2}}-\frac{1}{2}C_n,\\
  C_m&=&\sum_{j=1}^{m-1} \cos^{2p}{\left(\frac{\pi j}{m} \right)}\cos{\left( \frac{2r\pi j}{m}\right)} \nonumber.
\end{eqnarray}
A completely analogous procedure as for the even bipartition in this case leads to
\begin{equation}
  1+C_m=
 \begin{cases}
    0 &,\hbox{for }p=1,\ldots,r-1,\\
  \frac{m}{2^{2p}} {2p \choose p-r}&, \hbox{for }p=r,\ldots,m-r-1,
  \end{cases}
\end{equation}
and, as a consequence,
\begin{equation}
  h_{2r}(2p)=-\frac{1}{2},\quad \hbox{for }p=1,2,\ldots,\frac{n}{2}-r-1,
\end{equation}
which proves Eq.(\ref{eq:klih}).

We have therefore proved that $f_k(p)$ is for even $k$ zero at odd $1 \le p \le n-k-3$ (\ref{eq:ksod}), while for odd $k$ it is zero for even $2 \le p \le n-k-3$ (\ref{eq:klih}). For every other power $p$ the sum is nonzero. We have shown that such alternating sums of products of powers of roots of unity have rather magic properties.

\begin{figure}
\centerline{\includegraphics[width=.8\columnwidth]{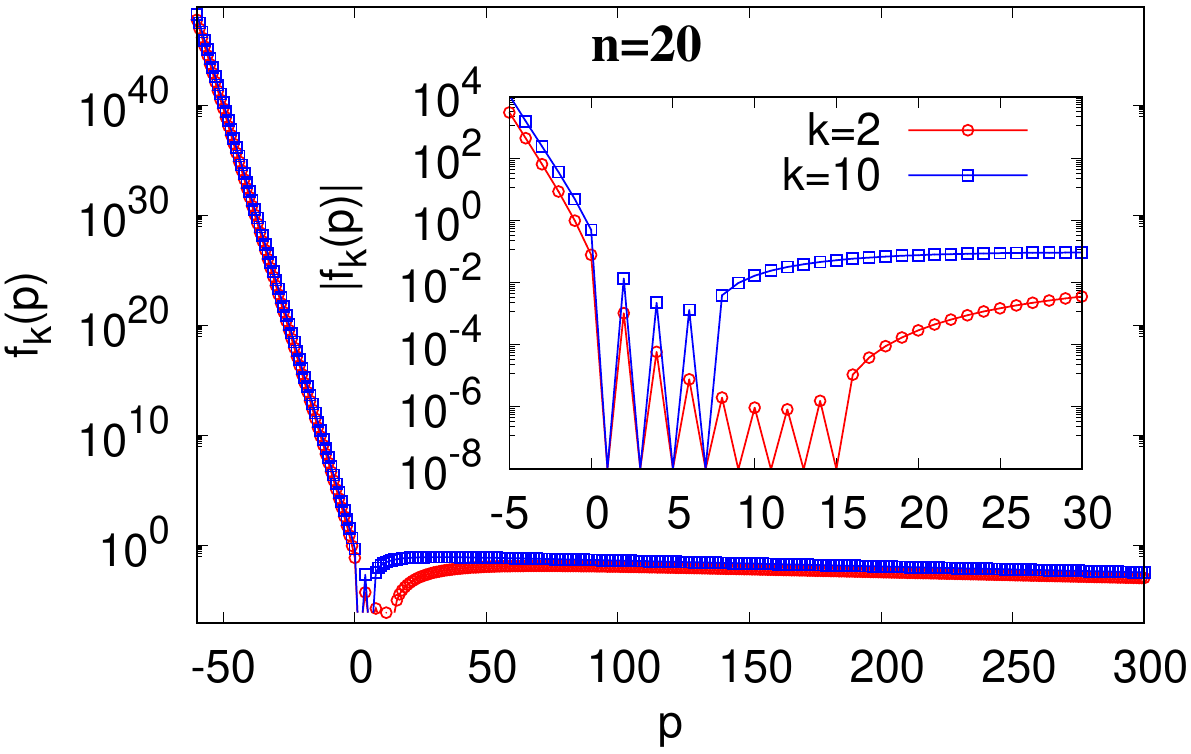}}
\caption{Values of $f_k(p)$ for $n=20$ and different $p$. In the inset one can see that $f_k(p)$ is zero (dips) for odd $p=1,3,\ldots,15$ at $k=2$, and for $p=1,3,5,7$ at $k=10$, as stated in Eq.(\ref{eq:ksod}).}
\label{fig:fkp}
\end{figure}

All this can be seen in Fig.~\ref{fig:fkp}. We can also see that $f_k(p)$ has very different behavior at negative $p$, where it is large and positive. This is exactly the regime that gives a diverging contribution to purity from $A_\lambda$, as discussed around Eq.(\ref{eq:rmin}), and is precisely canceled by the kernel contribution (Section~\ref{sec:kernel}).
In that context non-positive values of $p$ are also of interest. We are not going to study those in any detail, let us just state few values that will be used for the kernel contribution to purity at times $t=n/2-2$, and $t=n/2-3$, where the arguments are $p=-3,-2,-1,0$. One gets
\begin{equation}
  f_2(-1)=(-1)^{n/2},\quad f_2(-3)= (-1)^{n/2}\frac{n^2-4}{6},
  \label{eq:f2}
\end{equation}
holding for $n\ge 2$, then
\begin{equation}
  f_3(0)=-\frac{1}{2}(-1)^{n/2},\quad f_3(-2)= -(-1)^{n/2}\frac{n^2-28}{12},
\end{equation}
holding for $n\ge 6$, while $f_4(-1)=-2(-1)^{n/2}$ for $n\ge 6$, and $f_5(0)=\frac{1}{2}(-1)^{n/2}$ for $n\ge 8$.

\section{Scaling function for $k=2$}
\label{App:k2}

To express $f_2(p)$ in terms of known sums we first write $f_2(p)=h(p+3)-h(p+1)$, where $h(p)=2\sum_{j=1}^{n/2-1} (-1)^j \cos^p{\varphi_j}$. For large $p$ the high powers of $\cos{\varphi_j}$ will be very small, and one can approximate
\begin{equation}
  \cos^p{\varphi_j} \approx \exp{[-(\pi j/n)^2 p/2]}.
\end{equation}
On top of that we can extend the sum over $j$ in the definition of $h(p)$ to infinity because terms with $j \ge n/2$ are negligible for large $p$. This results in a sum that has a form of the Jacobi theta function (\ref{eq:th4}), more precisely\begin{equation}
h(p)=\vartheta_4(\cos^{p}{\varphi_1})-1.
\end{equation}
In the above expression we have also at the end replaced back a Gaussian approximation of the cosine with a cosine (otherwise one would not get the correct asymptotic limit of $\lambda_1$). We therefore have
\begin{eqnarray}
  f_2(p)&=&\vartheta_4(\cos^{p+3}{\varphi_1})-\vartheta_4(\cos^{p+1}{\varphi_1})\approx \nonumber \\
  &\approx& -\frac{\pi^2}{n^2} \cos^{p+1}{\varphi_1}\vartheta'_4(\cos^{p+1}{\varphi_1}).
\end{eqnarray}
Plugging that into Eq.(\ref{eq:leffp0}) gets us
\begin{eqnarray}
  \Delta \leff&=&(2\alpha)^2c^2\left[\frac{\vartheta'_4(c^{2t-n+6})}{\vartheta'_4(c^{2t-n+4})}-1 \right] \approx \\
  &\approx& -(2\alpha)^2 \left(\frac{\pi}{n}\right)^2 c^{2t-n+6} \frac{\vartheta''_4(c^{2t-n+4})}{\vartheta'_4(c^{2t-n+4})},
  \label{eq:razvojk}
\end{eqnarray}
where we abbreviated $c:=\cos{\varphi_1}=\cos{(\pi/n)}$, and the approximation leading to the 2nd line holds for large $n$ (due to Taylor expanding $\vartheta'_4(c^{2t-n+4}c^2)$ and the resulting $\cos^2{\varphi_1}-1$).

Derivatives can be also written out explicitly, resulting in
\begin{equation}
\Delta \leff(t)\approx -(2\alpha)^2 \left(\frac{\pi}{n}\right)^2 c^2 \frac{\sum_{j=1}^\infty (-1)^j j^2(j^2-1)c^{j^2 p}}{\sum_{j=1}^\infty (-1)^j j^2c^{j^2 p}},
\end{equation}
where $p:=2t-n+4$. This form is handy to get the asymptotic form of $\Delta \leff(t)$ on the scale $t \sim n^2$: taking the first nonzero term in the numerator, and likewise in the denominator, one gets
\begin{equation}
  \label{eq:k2lon}
  \Delta \leff(t) \asymp 12(2\alpha)^2 \left(\frac{\pi}{n}\right)^2 \left(\cos{\frac{\pi}{n}}\right)^{6t}.
\end{equation}
\begin{figure}
\centerline{\includegraphics[width=.86\columnwidth]{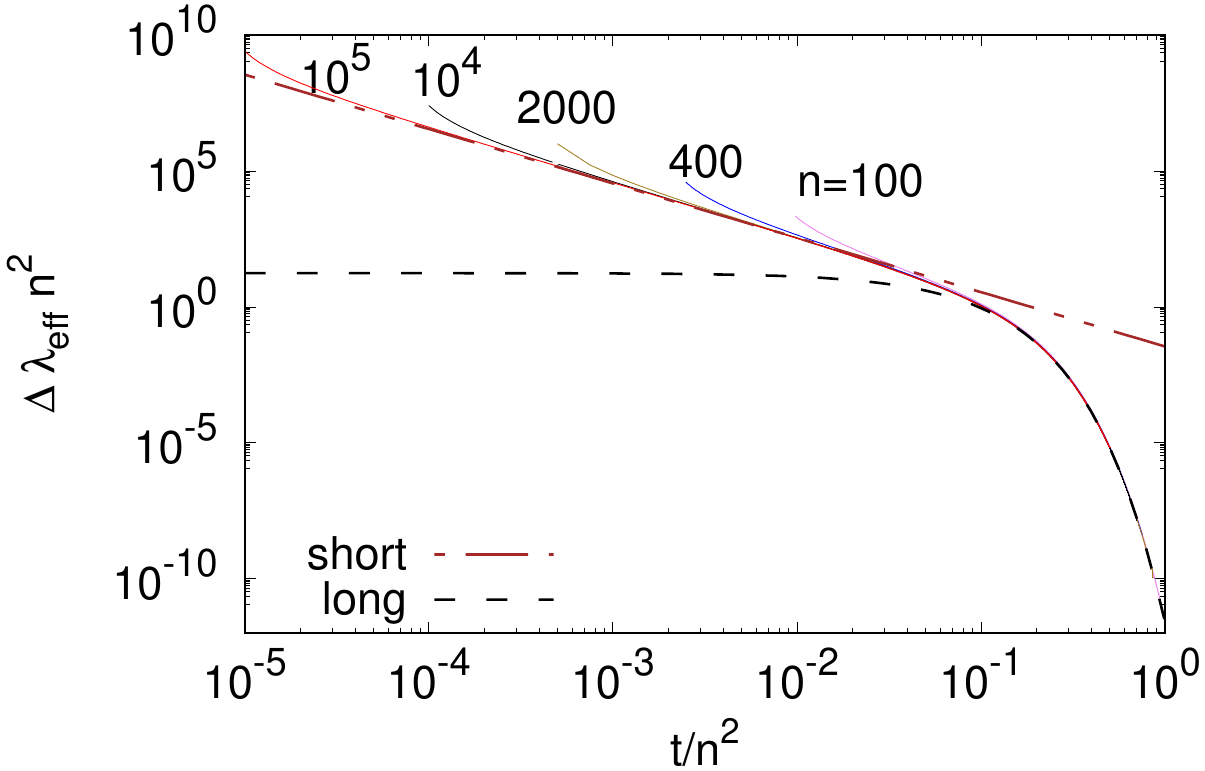}}
\caption{Long-time scaling of the decay rate $\Delta \leff$, where the scaling variable is $t/n^2$. Shown is expression in terms of theta functions, (\ref{eq:razvojk}), the long time exponential decay of Eq.(\ref{eq:k2lon}), and the short-time Eq.(\ref{eq:k2shor}), all without any fitting parameters. Bipartition $k=2$, $d=5$.}
\label{fig:k2lon}
\end{figure}
In Fig.~\ref{fig:k2lon} we can see that this long-time approximation works well from around $t/n^2 \gtrapprox 0.1$.

To get the short-time expansion we approach differently. What we need is expansion of $(\ln\vartheta'_4(q))'$, which we can write as $(\ln\vartheta'_4(q))'=(\vartheta''_4/\vartheta'_4-\vartheta'_4/\vartheta_4)+\vartheta'_4/\vartheta_4$. It turns out that for large $n$ and small $t$ (but larger than $n$) the first term is much smaller than the 2nd, and we keep only the second one. Using a product form of $\vartheta_4$
\begin{equation}
\vartheta_4(q)=\prod_{k=1}^\infty(1-q^{2k})\prod_{j=1}^\infty(1-q^{2j-1})^2,
\end{equation}
we can get by explicit derivatives
\begin{eqnarray}
  \label{eq:q}
  -\frac{\vartheta'_4(q)}{\vartheta_4(q)}&=&\sum_{j=1}^\infty \frac{2jq^{2j-1}}{1-q^{2j}}+\frac{2(2j-1)q^{2j-1}}{1-q^{2j-1}}=\\
  &=&\frac{\psi_{q^2}^{(1)}(1)}{2q\ln^2{q}}+\frac{\psi_q^{(1)}(1)-\psi_q^{(1)}(1-\ii \pi/\ln{q})}{\ln^2{q}} \nonumber
\end{eqnarray}
where $\psi_q^{(1)}(z)={\rm d}\psi_q(z)/{\rm d}z$ and $\psi_q(z)$ is a q-digamma function, specifically
\begin{equation}
  \psi_q^{(1)}(z)=\ln^2{q}\sum_{j=1}^{\infty}\frac{q^{j+z}}{1-q^{j+z}}+\frac{q^{2j+2z}}{(1-q^{j+z})^2}.
  \label{eq:psi1}
\end{equation}
Remembering that we need Eq.(\ref{eq:q}) for $q=(\cos{\frac{\pi}{n}})^{2t-n+4}$, which is for small $t$ close to $1$, we expand denominators in Eq.(\ref{eq:psi1}), finally getting
\begin{equation}
  \Delta \leff(t) \approx (2\alpha)^2 \frac{1}{4} \left( \frac{n}{t}\right)^2.
  \label{eq:k2shor}
\end{equation}
\begin{figure}
\centerline{\includegraphics[width=.85\columnwidth]{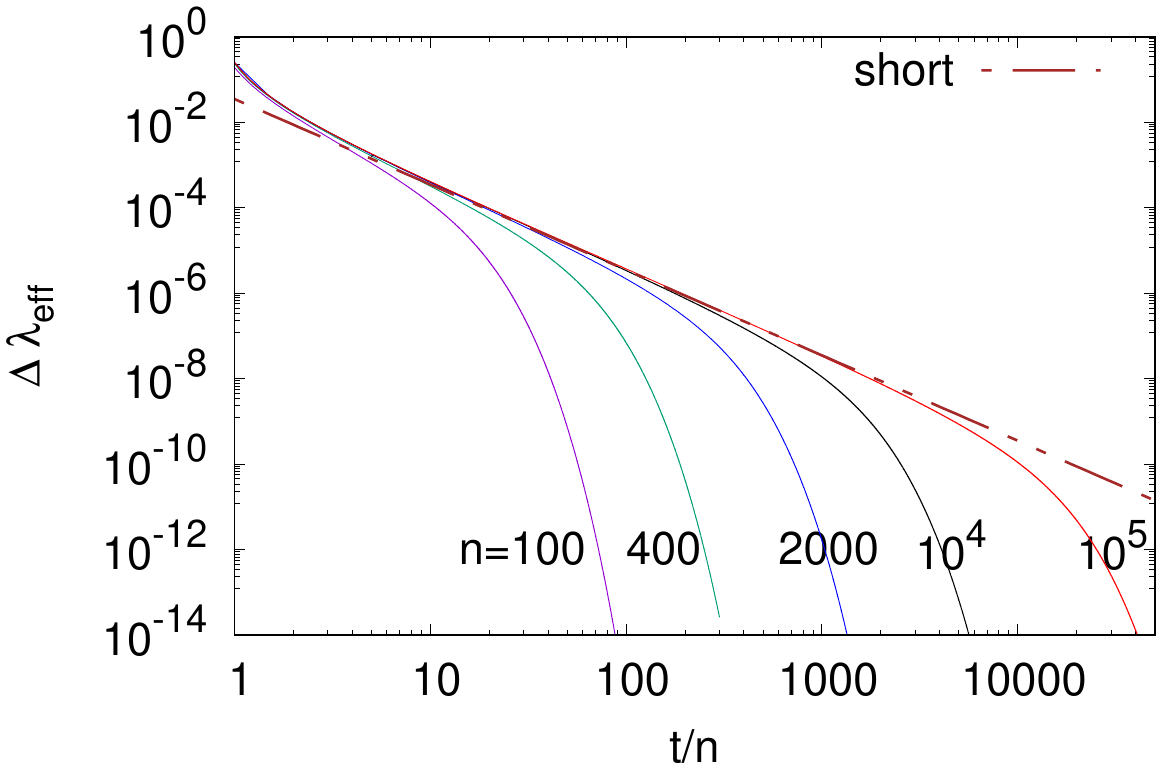}}
\caption{Short-time scaling of the decay rate $\Delta \leff$, where the scaling variable is $t/n$. Shown is expression in terms of theta functions, (\ref{eq:razvojk}), and the short time power-law of Eq.(\ref{eq:k2shor}) without any fitting parameters. The power law holds from about $t \approx 5n$. Bipartition $k=2$, $d=5$.}
\label{fig:k2shor}
\end{figure}
In Fig.~\ref{fig:k2shor} we can see the short-time approximation holds up to larger values of $t/n$ as one increases $n$. One can estimate that it is a good approximation from $t/n \approx 5$ until almost the time when the long-time approximation starts to hold. 

Note that derivatives of theta function that we use (\ref{eq:razvojk}) are with respect to $q$ and not with respect to the more usual $z$. However, one can also express everything in terms of derivatives with respect to $z$,
\begin{equation}
  \vartheta^{(1)}_4(q,z):={\rm d}\vartheta_4(q,z)/{\rm d}z.
\end{equation}
Namely, every theta function satisfies ``diffusion'' equation with $q$ being an imaginary time,
\begin{equation}
  -\frac{1}{4q} \vartheta^{(2)}_4=\vartheta'_4.
\end{equation}
Using that Eq.(\ref{eq:razvojk}) can be equivalently written as
\begin{equation}
\Delta \leff(t)=(2\alpha)^2 \left(\frac{\pi}{n}\right)^2 c^{2} \left( 1+\frac{\vartheta^{(4)}_4(c^{2t-n+4})}{\vartheta^{(2)}_4(c^{2t-n+4})}\right).
\end{equation}

\section{Scaling function for $k=n/2$}
\label{App:kn2}

\begin{figure}[ht]
\centerline{\includegraphics[width=.6\columnwidth]{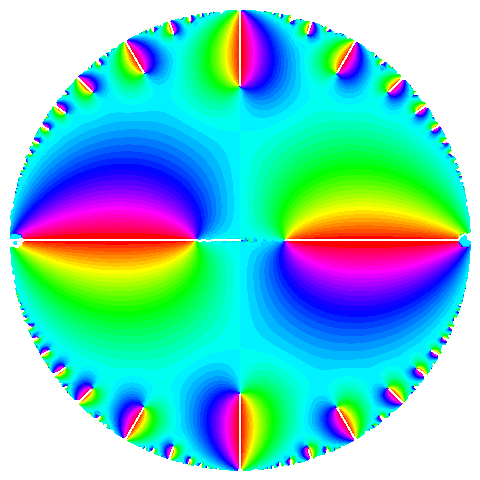}\hskip10pt\includegraphics[width=.1\columnwidth]{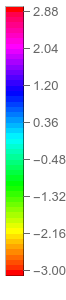}}
\caption{The phase of $q \vartheta'_1(z,q)/\vartheta_1(z,q)$ within a unit circle of complex $|q|<1$ at $z=\frac{\pi}{1000}$. Transition $\Delta \leff$ for bipartition with $k=n/2$ is given by such expression (\ref{eq:razth11}) evaluated at real values of $q \in [0,1]$ (visible as white line in figure).}
\label{fig:th1}
\end{figure}

We first observe that in the sum (\ref{eq:fkn2}) only terms with an odd $j=2k+1$ are nonzero, allowing us to write $f_{n/2}(p)=\sum_{k=0}^{n/4-1} (-1)^k (\cos{\varphi_{2k+1}})^p \sin{\varphi_{2k+1}}$. Again approximating high power of the cosine with a Gaussian, and letting the sum run to infinity, we get a sum that has a form of the Jacobi theta function~\cite{Abram} $\vartheta_1(z,q)$ (\ref{eq:th1}), giving
\begin{equation}
  f_{n/2}(p)\approx \frac{1}{2}\vartheta_1(\frac{\pi}{n},(\cos{\frac{\pi}{n}})^{4p}).
\end{equation}
Using Eq.(\ref{eq:leffp0}) and expanding for large $n$ results in
\begin{eqnarray}
  \Delta \leff(t) &\approx& (2\alpha)^2 \left( \frac{\pi}{n}\right)^2 \left[1-4c^{x_0}\frac{\vartheta'_1(\frac{\pi}{n},c^{x_0})}{\vartheta_1(\frac{\pi}{n},c^{x_0})} \right]=\nonumber \\
  &=& (2\alpha)^2 \left( \frac{\pi}{n}\right)^2 \left[ 1+\frac{\vartheta^{(2)}_1(\frac{\pi}{n},c^{x_0})}{\vartheta_1(\frac{\pi}{n},c^{x_0})} \right].
  \label{eq:razth11}
\end{eqnarray}
with $x_0:=8t-2n+4$ and $c:=\cos{(\pi/n)}$, and in the 2nd line we also expressed it in terms of derivatives with respect to $z$, $\vartheta^{(2)}_1:={\rm d}^2 \vartheta_1/{\rm d}z^2$. See Fig.~\ref{fig:th1} for an illustration of the Jacobi theta function in complex $q$ plane.

By using a product form of $\vartheta_1$,
\begin{equation}
  \vartheta_1(z,q)=2q^{1/4}\sin{z}\prod_{k=1}^\infty(1-q^{2k})\prod_{j=1}^\infty(1-2q^{2j}\cos{2z}+q^{4j}),
\end{equation}
we can explicitly calculate the required logarithmic derivative, obtaining
\begin{equation}
  \left[ 1+\frac{\vartheta^{(2)}_1(z,q)}{\vartheta_1(z,q)} \right]=\sum_{j=1}^\infty \frac{16 q^{2j}}{(1-q^{2j})^2}\cos{(2zj)}+\frac{8jq^{2j}}{1-q^{2j}}.
  \label{eq:th1sum}
\end{equation}
The benefit if this expression, compared to just plugging definition $\vartheta_1$ (\ref{eq:th1}) into Eq.(\ref{eq:razth11}), is that we have a single sum instead of a ratio of two sums, making extracting the asymptotic behavior simpler. Namely, for $t \sim n^2$ the argument $q$ is very small, and the whole sum can be approximated by its first term, which is $\approx 24q^2$, resulting in the long-time approximation
\begin{equation}
  \Delta \leff(t) \asymp 24(2\alpha)^2 \left( \frac{\pi}{n}\right)^2 \left( \cos{\frac{\pi}{n}}\right)^{16t}.
  \label{eq:kn2lon}
\end{equation}

\begin{figure}[t!]
  \centerline{\includegraphics[width=.83\columnwidth]{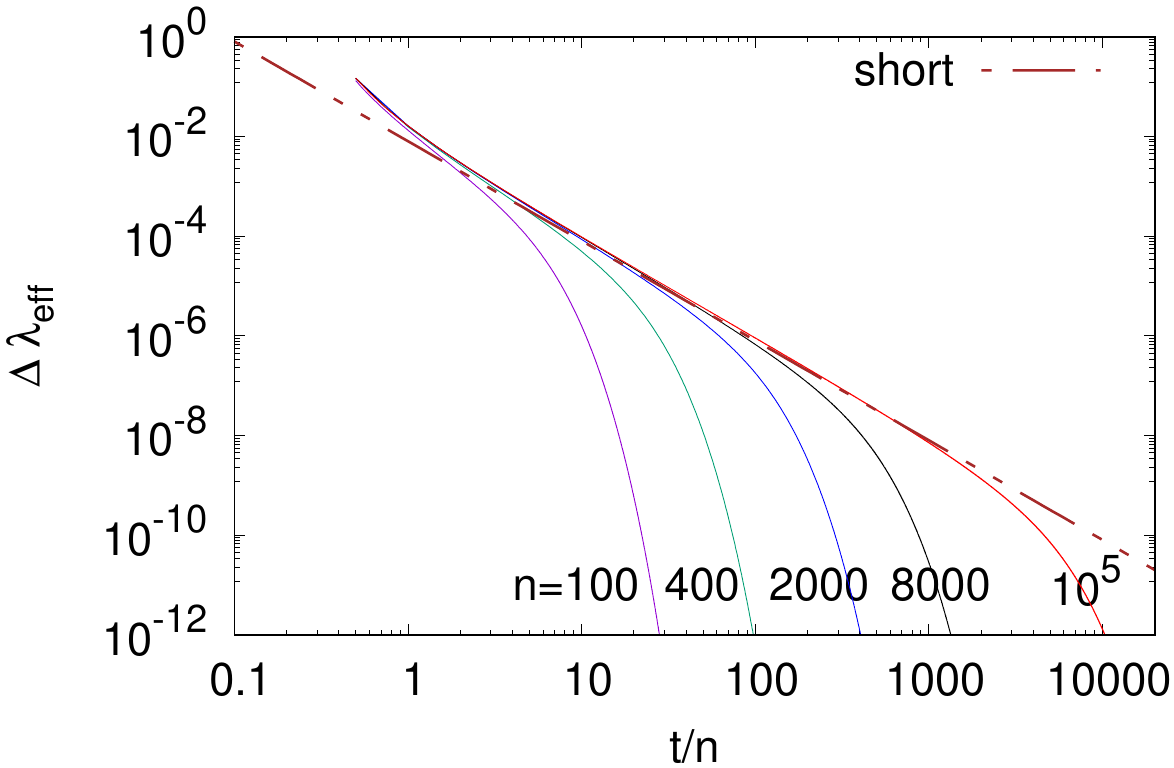}}
  \centerline{\includegraphics[width=.83\columnwidth]{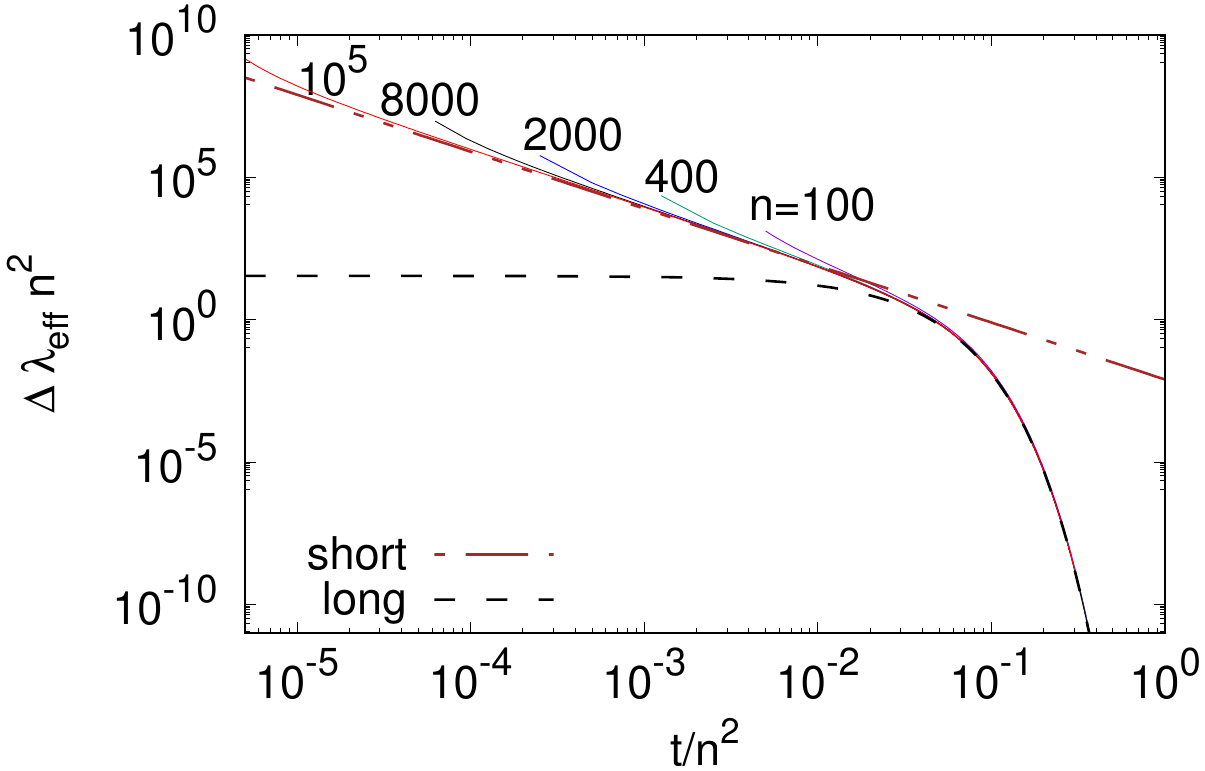}}
  \caption{Scaling functions for $k=n/2$, and $d=4$. Top: Short-time scaling of the decay rate $\Delta \leff$, where the scaling variable is $t/n$. Shown is expression in terms of theta functions, Eq.~(\ref{eq:razth1}), and the short time power-law of Eq.(\ref{eq:kn2shor}). The power law holds from about $t \approx 2n$ until increasingly larger $t/n\sim 0.01n$ as one increases $n$. Bottom: Long-time scaling where the scaling variable is $t/n^2$. Shown is expression in terms of theta functions, Eq.~(\ref{eq:razth1}), and the long time exponential decay of Eq.(\ref{eq:kn2lon}).}
  \label{fig:kn2lss}
\end{figure}
At short times, that is on a scale $t \sim n$, $q$ is on the other hand close to $1$, and one can expand $1-q^{2j}\approx (\pi/n)^2 8tj$. The 2nd sum in Eq.(\ref{eq:th1sum}) turns into a simple geometric sum, evaluating to $q^{2}n^4/(8\pi^4 t^2)\approx n^4/(8\pi^4 t^2)$, while the 1st sum is of type $\sum_j q^{2j}/j^2$, summing to $n^4/(4\pi^4 t^2) {\rm Li}_2(q^2)\approx n^4/(4\pi^4 t^2) \pi^2/6$, where ${\rm Li}_n(z)=\sum_{j=1}^\infty z^j/j^n$ is a polylogarithm. The specific dilogarithm is ${\rm Li}_2(1)=\zeta(2)=\pi^2/6$, giving short-time form
\begin{equation}
  \Delta \leff(t)\approx (2\alpha)^2 \left(\frac{1}{24}+\frac{1}{8\pi^2} \right)\left( \frac{n}{t}\right)^2.
  \label{eq:kn2shor}
\end{equation}
We again see that the short time scaling variable is $t/n$ (\ref{eq:kn2shor}), while the long time scaling variable is $t/n^2$ (\ref{eq:kn2lon}). Fig.~\ref{fig:kn2lss} demonstrated that both short and long time approximations work well.


\begin{thebibliography}{99}
\bibitem{footX} If one would have some external noise, like e.g. measurements, the purity could decrease even without any partial trace. Such situation is not studied in the present work.
\bibitem{foot1} We shall use a bold-face notation $\mathbf{x}$ for vectors, $[\mathbf{x}]_k$ for the $k$-th component in a given basis, and $\ket{\mathbf{x}}$ for a basis-independent notation.
\bibitem{foot1a} Note that the correct thermodynamic limit means one needs to first take the system size to $\infty$, only then $t \to \infty$.
\bibitem{foot2} The last component of vector $\mathbf{I}$ corresponds to $I_n(t)$ which is always $1$, $I_n(t)=1$, while the first component corresponds to $I_0(t)$, which is likewise $I_0(t)=1$. Purity $I_1(t)$ on the other hand does not couple back to $I_k(t)$ with $k=2,\ldots,n-1$, and can be obtained from a standalone recursion $I_1(t+1)=\alpha+\alpha I_2(t)$~\cite{PRR}.
\bibitem{foot4} Components of right steady state with odd symmetry are $[\R_0']_{n/2-k}=\alpha^{n/2-k} p_{2k}/p_{n}$ for $k=1,\ldots,n/2-2$, $[\R_0']_{n/2+k}=-\alpha^{n/2-k} p_{2k}/p_{n}$ for $k=1,\ldots,n/2-1$, and $[\R_0']_{n/2}=0$,$[\R_0']_{1}=1$, $[\R_0']_{n}=-1$, and where $p_n=\sum_{k=0}^{\lceil \frac{n}{4} \rceil-1}(-1)^k \alpha^{2k} {\frac{n}{2}-k-1 \choose k}$.
\bibitem{foot3} In order to get accurate numerical spectra for small $\varepsilon$ one should avoid using numerical diagonalization with floating-point numbers. What we do is first write matrix elements of $T(\varepsilon)$ as rational numbers (with huge numerators and denominators to be accurate with precision much better than $\varepsilon$), and then calculate the spectrum for this rational matrix via a characteristic polynomial. While such calculation becomes costly for larger $n$, it is still feasible for e.g. $n=20$ or $40$.
\end{thebibliography}
\end{document}